\documentclass[twocolumn,showpacs,preprintnumbers,prb,floatfix]{revtex4}
\usepackage{epsfig,amssymb,amsmath,hyperref,bm}
\usepackage{epsfig,amssymb}
\usepackage{graphics}
\usepackage{graphicx}
\usepackage{psfrag}
\newcommand\beq{\begin{equation}}
\newcommand\eeq{\end{equation}}
\newcommand\bea{\begin{eqnarray}}
\newcommand\eea{\end{eqnarray}}
\newcommand{\nonum}{\nonumber}

\newcommand\on{{\bar \omega}_n}

\begin{document}

\title{Transport through constricted quantum Hall edge systems: beyond
the quantum point contact}

\author{Siddhartha Lal}
\affiliation{The Abdus Salam ICTP, Strada Costiera 11, Trieste 34014, Italy.}
\email{slal@ictp.it}

\begin{abstract}
Motivated by surprises in recent experimental findings, we study transport in
a model of a quantum Hall edge system with a gate-voltage controlled constriction.
A finite backscattered current at finite edge-bias is explained from a
Landauer-Buttiker analysis as arising from the splitting of edge current
caused by the difference in the filling fractions of the bulk ($\nu_{1}$)
and constriction ($\nu_{2}$) quantum Hall fluid regions.
We develop a hydrodynamic theory for bosonic edge modes inspired by this model.
The constriction region splits the incident long-wavelength chiral edge
density-wave excitations among the transmitting and reflecting edge states
encircling it. The competition between two interedge tunneling processes taking
place inside the constriction, related by a quasiparticle-quasihole (qp-qh)
symmetry, is accounted for by computing the boundary theories of the system.
This competition is found to determine the strong coupling configuration of the
system.  A separatrix of qp-qh symmetric gapless critical states is found to
lie between the relevant RG flows to a metallic and an insulating configuration
of the constriction system. This constitutes an interesting generalisation of
the Kane-Fisher quantum impurity model. The features of the RG phase diagram
are also confirmed by computing various correlators and chiral linear
conductances of the system. In this way, our results find excellent agreement
with many recent puzzling experimental results for the cases of $\nu_{1}=1/3,~1$.
We also discuss and make predictions for the case of a constriction system with
$\nu_{2}=5/2$.
\end{abstract}

\pacs{73.23.-b, 71.10.Pm, 73.43.Jn}


\maketitle

\section{Introduction}
Despite being a subject of intense experimental and theoretical
interest, much is yet to be learnt of the combined effects of electron
correlations and impurities on the transport properties of
low-dimensional strongly correlated systems. The availability of several
non-perturbative theoretical methods for studying the physics of systems
in one spatial dimension has, however, allowed for considerable progress
to be made for such systems \cite{giamarchi}.
A physical one-dimensional system ideal for studying these issues
are fractional quantum Hall edges (FQHE) \cite{wen,wenint}. Considerable
experimental advances have been made in exploring the physics
of the edge states \cite{changrev} and in confirming many of the
theoretical predictions made of the remarkable properties of these
systems \cite{wenrev}. Several recent experiments have, however,
pointed out the need to develop a deeper theoretical understanding
of inter-edge quasi-particle tunneling phenomena in FQHE systems
with gate-voltage controlled constrictions
\cite{roddaro1,roddaro2,chung,comforti}. These experiments serve as the
primary motivation for the models proposed in this work.
However, before discussing these experiments, we first present a
discussion of the existing theoretical paradigm for the understanding
of inter-edge tunneling physics in FQHE systems.
\par
Kane and Fisher \cite{kane,fisher} observed in their classic work
that (a) the tunneling between two FQH edges separated by the FQH
fluid was akin to the backscattering of electrons by an impurity in
a Tomonaga-Luttinger liquid (TLL)
and (b) the tunneling between two FQH bubbles separated by vacuum was akin
to the tunneling of electrons across a weak-link (infinitely high barrier)
between two TLLs.
Their perturbative analysis revealed that process (a) was relevant
under RG transformations while process (b) was irrelevant, thus
suggesting that the low-energy physics of the FQHE tunneling problem
was likely to be that of two FQH bubbles separated by vacuum. Both
these scenarios are described by the boundary sine-Gordon model
\cite{gogolin}. In the following years, a quantum Monte-Carlo
simulation by Moon etal. \cite{moon}, an instanton calculation by
Furusaki and Nagaosa \cite{furusaki}, a conformal field-theory
analysis by Wong and Affleck \cite{wong} as well as the exact
solution of Fendley etal. \cite{fendley} using the thermodynamic
Bethe Ansatz method demonstrated that these scenarios were,
in fact, correct in their description of the system. Further, they
showed that, within the confines of the boundary sine-Gordon model,
there was no intermediate fixed point in the RG flow of the
backscattering/tunneling couplings in this model. Several works have
also analysed the effects of inter-edge interactions
\cite{oreg,pryadko,apalkov}, disorder \cite{kane2} on quasiparticle transport
in FQH edge systems. Attention has also been given to tunneling at
point-contacts between FQH fluids with different
filling-fraction \cite{chklovskii} as well as at contacts with Fermi liquid
reservoirs~\cite{chamonfrad}. More recently, attempts have been made at
developing a more general theory for the study of critical points
in edge tunneling between generic FQH states \cite{moore1,moore2}.
\par
\begin{figure}[htb]
\begin{center}
\scalebox{0.5}{
\psfrag{1}[bl][bl][2][0]{$\phi_{1,in}$}
\psfrag{2}[bl][bl][2][0]{$\phi_{1,out}$}
\psfrag{3}[bl][bl][2][0]{$\phi_{2,in}$}
\psfrag{4}[bl][bl][2][0]{$\phi_{2,out}$}
\psfrag{9}[c][c][2][0]{$(x\sim 0)$}
\includegraphics{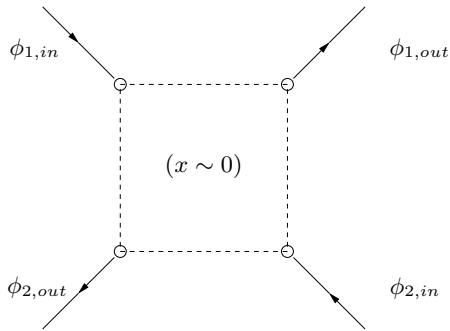}
}
\end{center}
\caption{A schematic diagram of the ``boundary" in our system given
by the dashed box around the region symbolised by $(x\sim 0)$. The
four chiral fields approaching and leaving this region are shown by
the arrows marked as $\phi_{1,in}$, $\phi_{1,out}$, $\phi_{2in}$ and
$\phi_{2,out}$. The dashed horizontal and vertical lines at the
junction represent quasiparticle transmission in various directions.}
\label{beam1}
\end{figure}

The phenomenological description of tunneling between chiral edges outlined
above relies on the following scenario. For no backscattering coupling
between the two edges of opposite chirality at, say, $x\sim 0$, we
have a system of two chiral 1D systems which are continuous
at $x=0$. This can be seen by consulting Figure (\ref{beam1}) given below
for the case of the fields $(\phi_{1,in}, \phi_{1out})$
and $(\phi_{2in}, \phi_{2out})$ being continuous.
Upon introducing a small RG-relevant inter-edge tunnel coupling,
we are left at strong backscattering coupling with a
system in which the earlier edges are now discontinuous across $x=0$;
they have, in fact, now become reconnected in a different configuration,
with the fields $(\phi_{1,in}, \phi_{2out})$ and
$(\phi_{2in}, \phi_{1out})$ now being continuous (as can be seen in
Figure (\ref{beam1}) below). This means that, in order to describe
ballistic transport intermediate between these two which is characterised by
a finite backscattering of current, one must consider the
possibility of the fields describing the chiral edge excitations as being
discontinuous across $x=0$. In doing so, it appears necessary to rely on
ideas non-perturbative in nature. Accounting for additional
quasiparticle tunneling among the various incoming and outgoing
chiral edges is then likely to lead to a non-trivial variation of
the boundary sine-Gordon model. Insights on these issues were gained
recently in Ref.(\cite{epl}), in the form of a new model for the constriction
geometry in quantum Hall system which, while being simple in essence, is
clearly beyond the paradigm of the quantum point contact. We aim here to
develop the ideas presented in that work, exploring more fully the
consequences of such a constriction system.
\par
As will be discussed in the next section, several recent experiments on
inter-edge tunneling in FQHE systems show that it is possible to use
the voltage of a split-gate constriction to tune the inter-edge
transmission to values intermediate to those in the two scenarios
described above. Further, they reveal a very interesting evolution
of the transmission through the constriction with decreasing
inter-edge bias. This will lead us to formulate a simple
phenomenological model for the split-gate constriction region. We will
then perform a Landauer-Buttiker analysis and compute the conductances
of the model. The results of this analysis will be seen to point to
some interesting conclusions for transport in the presence of a
constriction.
It is now well established that the low-energy theory for the
dynamics of the gapless long-wavelength excitations on the edges of a
FQH system are described by a hydrodynamic continuum chiral TLL
theory \cite{wen} of propagating density disturbances which are bosonic
in nature.
Adhering to the spirit of such a hydrodynamic description, we
formulate a continuum model for the constricted quantum Hall edge
system in section III.
In section IV, we introduce local quasiparticle tunneling processes inside the
constriction and construct a boundary theory for the problem. In this way, we
investigate the RG phase diagram of the system for the various tunnel couplings.
We complete the study in section V by computing several chiral correlators and
conductances at weak- and strong-quasiparticle tunnel coupling values.
We then present a comparison of the results of our model with those obtained
from recent experiments in section VI. Here, we will also reflect on the
relevance of our model to the case of a constriction with a filling
factor of $\nu=5/2$. We end by discussing some finer aspects of the model
and outlining some open directions in section VII.

\section{Model for a split-gate constriction}
We now propose a simple, phenomenological model for a
split-gate constriction created in a quantum Hall system.
A schematic diagram of an experimental setup of a FQH bar with a
gate-voltage controlled constriction
is shown below in Figure (\ref{kf1}).
\begin{figure}[htb]
\begin{center}
\scalebox{0.5}{
\psfrag{1}[bl][bl][3][0]{1}
\psfrag{2}[bl][bl][3][0]{2}
\psfrag{3}[bl][bl][3][0]{3}
\psfrag{4}[bl][bl][3][0]{4}
\psfrag{5}[bl][bl][3][0]{G}
\psfrag{6}[bl][bl][3][0]{G}
\psfrag{7}[bl][bl][3][0]{S}
\psfrag{8}[bl][bl][3][0]{D}
\psfrag{9}[bl][bl][3][0]{$\nu_{1}$}
\psfrag{10}[bl][bl][3][0]{$\nu_{1}$}
\psfrag{11}[c][c][2.5][0]{$\nu_{2}$}
\includegraphics{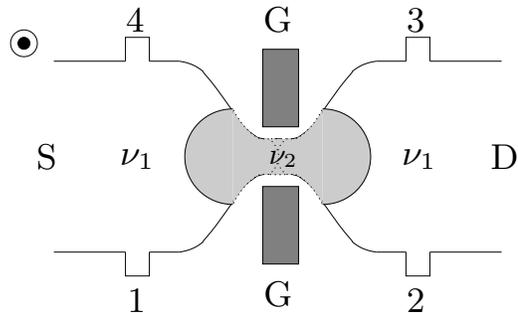}}
\end{center}
\caption{A schematic diagram of a FQH bar with a gate-voltage (G)
controlled split-gate constriction which lowers the electronic
density in the constriction region as well as brings the top and bottom
edges of the Hall fluid in close proximity, allowing for tunneling
to take place between
the opposite edges. S and D signify the source and drain ends
of the Hall bar while the numbers 1 to 4 signify the current/voltage
lead connections. The external magnetic field points out of the plane
of the paper.}
\label{kf1}
\end{figure}
\par
As is indicated in Figure
(\ref{kf1}), the constriction is created electrostatically
in a two-dimensional electron gas (2DEG) quantum Hall system at
filling-fraction $\nu_{1}$ by the electronegative gating of
metallic split-gates. An important effect of the split-gate constriction
is to bring the two counter-propagating edges of the Hall fluid in close
proximity, allowing for the possibility for quasiparticles to tunnel
between them. As discussed earlier, this has been a major focus in
the study of the physics of FQHE systems. However, an often neglected
effect of the split-gates
is that the electric field induced by them reduces the 2DEG
density (and hence the filling-fraction of the Hall fluid) in the
narrow constriction region; the inter-particle
correlations in the constriction are thus likely to increase in
strength. We can, therefore, expect the filling-fraction
of the FQH fluid in the constriction,
$\nu_{2}$, to be a function of $\nu_{1}$ as well as the
gate-voltage $V_{g}$, i.e., $\nu_{2}\equiv\nu_{2}(\nu_{1},V_{g})$,
in such a way that (i) $\nu_{2}=\nu_{1}$ for
$V_{g}=0$ (i.e., no constriction) and (ii) $\nu_{2}<\nu_{1}$
for $V_{g}<0$ (i.e., with a constriction).
While the filling-factor $\nu_{1}$ (for $\nu_{1}^{-1}$ being an odd
integer, such that we have only single edge states) can be related
to the strength
of the inter-edge density-density interactions, $g_{edge}$, in the
bulk of the FQH system \cite{wenint,fisher,chklovskii}
\beq
\nu_{1} = (1 + g_{edge})^{-1/2}~,
\eeq
no such simple relation exists, at present, for the
filling-factor in the constriction, $\nu_{2}$. Clearly, this will need
a greater understanding of the role of the gate-voltage $V_{g}$ in
creating the constriction.

\subsection{Surprises from the experiments}
We now turn to a discussion of the several puzzling results observed in
experiments on transport through split-gate constrictions in
integer \cite{roddaro1} and fractional \cite{roddaro2,chung} quantum
Hall systems and outline the several intriguing results
observed therein. Working with an experimental setup as shown
in Fig.(\ref{kf1}), a finite dc bias between the two edges coming
towards the constriction $V_{c}$ is imposed through the source
(S) terminal while the drain (D) terminal as well as terminals $1$ and
$2$ are kept grounded.
\par
\noindent
(i) A current $I$ is incident on the constriction
from the upper-left edge and is partially transmitted with the
transmitted current finally being collected at the terminal 3. The reflected
current is collected at terminal 1 and gives rise to a
bias-independent longitudinal differential resistance across
the constriction at large bias $V_{c}$.
\par
\noindent
(ii) The two-terminal
differential conductance $G(V_{c})$
is measured at temperatures as low as $250 {\mathrm mK} < eV_{c}$
and gives the transmission coefficient of the constriction
$0 \leq t(V_{c})~(=G(V_{c})/G_{0}) \leq 1$ (where
$G_{0}=\nu_{1}e^{2}/h$ is the Hall conductance of the bulk; $\nu_{1}=1$ in
Ref.\cite{roddaro1} and $\nu=1/3$ in Ref.\cite{roddaro2,chung}). At
sufficiently large values of the gate-voltage $V_{g}$ and large bias
$V_{c}$, $t(V_{c})$ is observed to saturate with $|V_{c}|$ at a value less
than unity. Further, $t(V_{c})$ is observed to dip sharply and vanish with a
power-law dependence on $V_{c}$ as $|V_{c}|\rightarrow 0$. A comparison
with the theory of inter-edge Laughlin quasiparticle tunneling
developed by Fendley etal.
\cite{fendley} suggests strongly that the constriction transmission is
governed by the local filling-factor of the Hall fluid in the constriction,
even though this region is likely to be small in extent.
This is unexpected for the case of the bulk being in an
integer quantum Hall state \cite{roddaro1} where edge transport is
understood in terms of noninteracting electron charge carriers.
\par
\noindent
(iii) A particularly intriguing observation is that of
the evolution of the
constriction transmission $t(V_{c})$ at very low temperatures
(e.g., $50{\mathrm mK}$) as the split-gate voltage
$V_{g}$ is varied in the limit of vanishing inter-edge bias $V_{c}$.
While $t(V_{c})$ shows a
zero-bias minimum at sufficiently large $V_{g}$, decreasing $V_{g}$
leads to a bias-independent transmission at a particular value of $V_{g}$
and then to an enhanced zero-bias transmission for yet lower values
of $V_{g}$. The same behaviour of the zero-bias transmission is also
observed by holding the gate-voltage $V_{g}$ fixed and lowering the
temperature from $700{\mathrm mK}$ to $50 {\mathrm mK}$. For the case
of $\nu_{1}=1$, the bias-independent transmission is observed at a
value of $t^{*}=1/2$~~\cite{roddaro1} while for $\nu_{1}=1/3$, it is
observed at $t^{*}=3/4$~~\cite{roddaro2,chung}. A similar enhancement of
the zero-bias transmission at sufficiently weak gate-voltages was also
reported for the case of bulk filling-fractions $\nu_{1}=2/5$ and $3/7$
\cite{chung}. The bias-independence as well as the enhancement of the
transmission $t(V_{c})$ is quite unexpected from the viewpoint of the
theoretical framework of edge tunneling described earlier.
\par
\noindent
(iv) The constriction transmission for a bulk $\nu=2$ system displayed
two dip-to-peak evolutions, with bias independent behaviours observed
at $t^{*}=1/4, 1/2$ and $3/4$~~\cite{roddaro1}. This appears to indicate
the independent effects of the two edge modes in the $\nu=2$ system.
\par
\noindent
(v) Varying considerably the size and shape of the metallic gates
(which form the constriction region) did not appear to affect the
dip-to-peak nature of the evolution of the constriction transmission with
the strength of the gate voltage~~\cite{roddaro3}.
\par
Let us now consider the probable effects of a split-gate voltage constriction.
Clearly, other than promoting the tunneling of quasiparticles between
oppositely directed edges (due simply to enhanced wavefunction overlap due
to the proximity of the edges), the more noteworthy effect is likely to be the
creation of a
smooth and long constriction potential, which depletes the local electronic
density (and hence lowers the local filling factor) locally from its
value in the bulk. Indeed, this led Roddaro
and co-workers \cite{roddaro1} to conjecture on the possibility of a
small region in the neighbourhood of the constriction with a reduced
filling factor ($\nu_{2}<\nu_{1}$) as the cause of their puzzling
results (see fig.(\ref{kf1})). This conjecture, however, remained
unsubstantiated by the formal analysis of a concrete theoretical model.
Thus, their explanations for the $\nu=1$ system remained suggestive at best
and no attempts at unifying the observations at both integer and fractional
values of $\nu$ were made. Thus, the pressing questions that remain to be
answered are as follows. What drives the gate-voltage tuned insulator-metal transition at
vanishing edge-bias in the constriction system (as evidenced by the dip-to-peak
evolution with decreasing strength of the gate voltage)? Can purely local interedge
quasiparticle tunneling processes, which need an interplay of impurity scattering
and electronic correlations~~\cite{kane}, be the sole cause? Is there a symmetry
governing the edge-bias independent response of the constriction transmission at a
critical value of the constriction filling factor (as seen by tuning the gate voltage)?
If the system is indeed critical at this point, what does its gapless theory look like?
\par
At the same time, earlier theoretical efforts \cite{papastroh,papamac,agosta} were
unable to provide any simple explanations of these experimental observations. Most
notably, the scenario proposed in
ref.\cite{papastroh} involved the complications of stripe states arising
from longer range interactions. However, it failed to present any mechanism in
explaining the evolution of $g$ with $V_{G}$. The same is also
true of proposals of line junctions \cite{papamac} as well as the effects of
inter-edge interactions on quasiparticle tunneling \cite{agosta}.
Thus, keeping in mind that the theory of
refs.\cite{kane,fendley} matches the experiments in only a very restricted
parameter regime, the lack of a clear theoretical
understanding remained an important problem to be addressed. The creation of
a model with an effort towards explaining the puzzles was, therefore, the
main motivation of an earlier work~~\cite{epl}. In what follows, this model
is first formulated and then analysed in detail.

\subsection{Landauer-Buttiker analysis of transport}
Inspired by these experimental findings, we build, in the remainder of
this section, a simple phenomenological model
of a FQH split-gate constriction with a reduced local filling-fraction.
In this way, we aim to provide a qualitative understanding of some of the
observations discussed above. Furthermore, certain elements of this simple
model will then be employed as input parameters in a more sophisticated
theory involving bosonic edge excitations in subsequent sections
in providing explanations of some of the other, more puzzling,
experimental results. The analysis of this model will be carried out in
two ways. The first will involve an explicit calculation of the various
Landauer-Buttiker conductances of the measurement geometry. In the second
analysis, we will show how the results of the explicit calculation can
be derived more simply by making two assumptions of the system at hand.
\par
We begin by performing a Landauer-Buttiker analysis of the edge circuit
~~\cite{buttiker}. This is shown below in Fig.(\ref{circuit}).
\begin{figure}[htb]
\begin{center}
\scalebox{0.6}{
\psfrag{1}[bl][bl][3][0]{2}
\psfrag{2}[bl][bl][3][0]{3}
\psfrag{3}[bl][bl][3][0]{5}
\psfrag{4}[bl][bl][3][0]{6}
\psfrag{5}[bl][bl][3][0]{1}
\psfrag{6}[bl][bl][3][0]{4}
\psfrag{7}[bl][bl][2.5][0]{$I$}
\psfrag{8}[bl][bl][2.5][0]{$I_{tr}$}
\psfrag{9}[bl][bl][3][0]{$\nu_{1}$}
\psfrag{10}[bl][bl][3][0]{$\nu_{1}$}
\psfrag{11}[c][c][2.5][0]{$\nu_{2}$}
\psfrag{12}[bl][bl][2.5][0]{$I_{ref}$}
\includegraphics{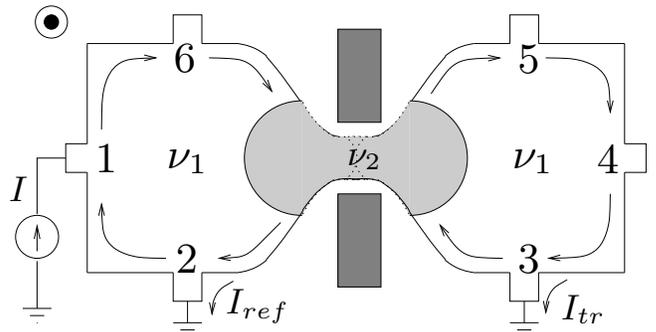}}
\end{center}
\caption{A schematic diagram of a QH bar constriction circuit with the
filling factors of the bulk and constriction regions being $\nu_{1}$
and $\nu_{2}$ respectively. The numbers 1 to 6 signify the various
current and voltage terminals. The current $I$ is sent into the circuit
at terminal 1 by the current source, while the current transmitted
through ($I_{tr}$) and the current backscattered from ($T_{ref}$) the
constriction are received at terminals $3$ and $2$ respectively. The
external magnetic field points out of the plane of the paper.}
\label{circuit}
\end{figure}
The central feature of our model is the region of lowered filling factor
($\nu_{2}$) assumed to be created by the split gate constriction gates. Let us
now estimate the spatial extent, $L_{con}$, of the $\nu_{2}$
region. This can be done by noting that the transport data taken at a temperature of
$50\mathrm{mK}$ does not appear to show any interference effects arising from
coherence across the entire constriction \cite{roddaro1}. Thus, $L_{con}$ can
be safely assumed to be longer than the thermal length $L_{T}=hv/k_{B}T$
(where $v$ is the edge velocity). For a typical $v=10^{3}\mathrm{ms}^{-1}$
~~\cite{komiyama} and $T=50\mathrm{mK}$, $L_{T}\sim 1\mu {\mathrm m}$. Clearly,
$L_{con}(>1 \mu {\mathrm m})>>$ magnetic length $l_{B}(\sim 100{\mathrm \AA{}})$,
justifying our assumption of the mesoscopic nature of the $\nu_{2}$
region.
\par
In a Landauer-Buttiker analysis~~\cite{buttiker1}, the net currents flowing in
the various arms are assumed to satisfy linear relations with the applied
voltages (valid for small values of the voltages), with the proportionality
factors being the various transmission coefficients for the quantum system.
Solving the various linear relations between the currents and voltages gives
us the various conductances of the system. It is helpful to use the fact that
the net current for voltage probes is zero, and that we have the freedom to set
the voltage of one of the terminals to zero (as currents are related to applied
voltage differences). Thus, in Fig.(\ref{circuit}), we set $V_{4}=0$, and since
terminals $4, 5$ and $6$ are voltage probes, $I_{4}=0=I_{5}=I_{6}$. When put together
with the fact that terminals 2 and 3 are grounded, i.e., $V_{2}=0=V_{3}$, this
allows us to exclude the current-voltage relation for terminal $4$ altogether
(i.e., remove one row from the transmission matrix linking the currents and
voltages). Thus, we can write the current-voltage relations in matrix form as
\beq
{\bf I} = \mathbf{{\bar T}}~ {\bf V}
\eeq
where the current and voltage column vectors are
${\bf I} = (I_{1},I_{2}, I_{3}, I_{5}, I_{6})$ and
${\bf V} = (V_{1}, V_{2}, V_{3}, V_{5}, V_{6})$ respectively and the
$\mathbf{{\bar T}}$ transmission matrix is given by
\begin{displaymath}
\mathbf{{\bar T}} =
\left(\begin{array}{ccccc}
\nu_{1} & -\nu_{1} & 0 & 0 & 0\\
0 & \nu_{1} & -\nu_{2} & 0 & -\nu_{ref}\\
0 & 0 & \nu_{1} & -\nu_{1} & 0\\
0 & 0 & -\nu_{ref} & \nu_{1} & -\nu_{2}\\
-\nu_{1} & 0 & 0 & 0 & \nu_{1}
\end{array} \right)~,
\end{displaymath}
where $\nu_{ref}$ is the transmission coefficient for the current backscattered
from the constriction. We now solve these linear relations.
Measuring all voltages with respect to terminal $4$ (which we have set to
zero), we can see that as $I_{6}=0$, we find $V_{6}=V_{1}$. Further, from
$I_{5}=0$, we get
\beq
V_{5} = \frac{\nu_{2}}{\nu_{1}} V_{6} = \frac{\nu_{2}}{\nu_{1}} V_{1}~.
\eeq
The current leaving the circuit at terminal 3 is given by
$I_{3} = -I_{tr} = -\nu_{1} V_{5}$
(where $I_{tr}$ is the current transmitted through the constriction region
from terminal $6$ to terminal $5$). This gives us
\beq
I_{tr} = \nu_{1}\frac{\nu_{2}}{\nu_{1}} V_{1} = \nu_{2} V_{1}~.
\eeq
In a similar manner, we can compute the current leaving the circuit at terminal
$2$ (which, with terminal $3$ being grounded, consists entirely of the current
backscattered from the constriction) as $I_{2} = -I_{ref} = -\nu_{ref} V_{6}$.
Then, from overall current conservation in our circuit, the total injected
current is given by $I_{1}=\nu_{1}V_{1}= I_{tr} + I_{ref}$, which gives us
\beq
I_{ref} = (\nu_{1} - \nu_{2}) V_{1}~.
\eeq
This leads us to $\nu_{ref} = \nu_{1}-\nu_{2}$. This expression for $\nu_{ref}$
can also be found very simply by noting that the constraint of unitarity for
the transmission matrix means that the sum of the elements in every row (or
every column) must add up to zero~~\cite{buttiker1}. We can now also compute
the conductance (in units of $e^{2}/h$) due to the current backscattered
from the constriction as
\beq
G^{back} = \frac{I_{ref}}{V_{1}} = \nu_{1}-\nu_{2}~.
\eeq
This also gives us the ``background" value of the resistance drop across the
constriction as
\beq
R^{BG} = \frac{V_{6} - V_{5}}{I_{1}-I_{5}} = \frac{G^{back}}{\nu_{1}^{2}}~.
\eeq
\par
Having carried out the Landauer-Buttiker analysis, we now show how all of
the results obtained therein can be rederived through
a simple analysis of the circuit which relies on essentially two assumptions
on the nature of the system at hand and the conservation of current~~\cite{epl}.
This will allow us to reflect on the simplicity and efficiency of the assumptions.
Thus, let us begin by stating the assumptions made and show
how they lead in a straightforward way to simple relations for several
physical quantities measured in the experiment. These are:\\
(i) the voltage bias between the two edges of the sample (i.e., the
Hall voltage for the system being in a quantum Hall state) is not affected
by the local application of a gate-voltage at a constriction as long
as the bulk of the system is in an incompressible quantum Hall state
with filling-fraction $\nu_{1}$,\\
(ii) the two-terminal conductance measured across the constriction is
determined by the current transmitted through it, which in turn is
governed by the filling-fraction of the Hall fluid in the constriction,
$\nu_{2}$. This needs the breakup of the current coming towards the
constriction to take place at the boundary and constriction Hall fluid
regions  (which is sufficiently far away from the center of the
constriction region).
\par
Thus, by denoting the current injected into the system from the source
terminal as $I$, we know that $I = G_{b} V_{63}$ where
$G_{b}=\nu_{1}e^{2}/h$ is the bulk Hall conductance and $V_{63}$ is
the edge-bias. From assumption (ii), denoting the current transmitted
through the constriction as $I_{tr}$, it is clear that
$I_{tr} = G_{c} V_{63}$, where $G_{c}=\nu_{2}e^{2}/h$ is the two-terminal
conductance measured across the constriction. Putting these two relations
together using assumption (i), we obtain
the transmitted current $I_{tr}$ in terms of the incoming current $I$ as
\beq
I_{tr} = \frac{G_{c}}{G_{b}} I = \frac{\nu_{2}}{\nu_{1}} I~.
\label{transcurr}
\eeq
Thus, we see that our assumptions give us a very simple relation for the
the splitting-ratio $\gamma$ for the currents at the constriction
(which is simply related to the transmission coefficient of the constriction
discussed above for no inter-edge tunneling) as being
$\gamma = \nu_{2}/\nu_{1}$.
Now, from Kirchoff's law for current conservation, we get the current
reflected at the constriction
$I_{ref} = I - I_{tr} = (1 - \nu_{2}/\nu_{1}) I$. This then gives the
minimum value of the backscattering conductance as
\beq
G^{back}=I_{ref}/V_{63} = (1 - \nu_{2}/\nu_{1}) G_{b}
= (\nu_{1} - \nu_{2})\frac{e^{2}}{h}~.
\label{backcond}
\eeq
$G_{back}$ is simply related to the reflection coefficient of the
constriction for no inter-edge tunneling, and shows that the {\it effective}
filling fraction governing $G_{back}$ is $\nu_{ref} = \nu_{1} - \nu_{2}$.
Now, with the current at terminal $5$, $I_{5}$, being
the transmitted current $I_{tr}$, we get
$I_{5} = G_{b}V_{5} = I_{tr} = G_{c} V_{6}$ (since $V_{3}=0$), giving
$V_{5} = (G_{c}/G_{b}) V_{6}$. We then find the ``background" value of
the longitudinal resistance drop across the constriction to be
\beq
R^{BG}=\frac{V_{6} - V_{5}}{I_{1}-I_{5}}
= (1 - \frac{\nu_{2}}{\nu_{1}}) G_{b}^{-1}
\label{bgroundres}
\eeq
which arises from the partial reflection and transmission of the
incoming edge current due to the mismatch of the filling-fraction in the bulk
and constriction regions. The experimentally obtained value for $R^{BG}$ is,
in fact, used by the authors of Refs.\cite{roddaro1,roddaro2} to
determine the value of the constriction filling-factor $\nu_{2}$
from eq.(\ref{bgroundres}). Further, we can see that $G^{back}$ and
$R^{BG}$ are simply related by $G^{back} = G_{b}^{2} R^{BG}$.
More generally, the differential longitudinal drop across the
constriction $dV_{65}/dI$ is related to (and also experimentally determined in
\cite{roddaro2}) the differential
backscattering conductance $dI_{ref}/dV_{63}$ by the simple relation, as seen
earlier
\beq
\frac{dI_{ref}}{dV_{63}} = G_{b}^{2} \frac{dV_{65}}{dI}~.
\label{restocond}
\eeq
Further, we also check that the Hall conductances measured on the two
sides of the constriction are determined by $\nu_{1}$ alone
\beq
\frac{I_{tr}}{V_{53}} = G_{b} = \frac{I}{V_{62}}~.
\eeq
Thus, we see that by allowing for the constriction region to have a
reduced filling-fraction ($\nu_{2}$) than that of the bulk ($\nu_{1}$)
and making the two assumptions stated above, we are able (i) to find
a simple expression for the splitting-ratio $\gamma$ of the current
incident on the constriction (or, the zeroth constriction transmission
coefficient) as well as (ii) find an expression for the longitudinal
resistance drop across the constriction which arises from the breakup
of the current.
\par
At the heart of these results lies the fact that a constriction
region with a reduced filling-fraction necessitates the transfer of charge
from the incoming edge to the opposite outgoing edge via the
incompressible bulk. Put another way, it becomes imperative to consider the
non-conservation of edge current in studying transport across
such a constriction. This is characterised by the presence of a
current reflected at the boundary of the bulk and constriction regions in
the model setup above. While charge dissipation away from the edge can be
modeled in terms of quasiparticle tunneling at multiple point-contact
junctions~\cite{chamonfrad,ponomarenko}, such a mechanism appears to be
incompatible with the experimental finding of an
edge bias-independent current reflected from the constriction
region. The existence of a narrow gapless region of Hall fluid lying in
between the incompressible bulk and constriction Hall fluid regions may
well provide an answer: such a gapless region would act as a
channel for the current reflected from the constriction region. It is,
therefore, tempting to speculate on the possibility
of a non-perturbative physical mechanism \cite{neto} of a chiral
Tomonaga-Luttinger liquid undergoing charge dissipation along a short
stretch of its length while in contact with a bath (the gapless region)
as being the microscopic origin for the phenomenological model
described above.
\par
While there are ways of studying the electrostatic effects of
a gate-voltage controlled constriction on the incompressible quantum Hall
fluid \cite{chklovskii,papastroh}, we have instead chosen a particularly simple and
tractable path for modeling the edge structure which involves very few
details pertaining to the bulk. The electrostatic calculations of Ref.
\cite{papastroh} explore the possibility of edge reconstruction within the
constriction region, i.e., long-range interactions between
electrons in the quantum Hall ground state giving rise to a set of
compressible and incompressible stripes at the edge \cite{yoshioka}. In this
work, however, we consider only short ranged electron correlations, which
cause the formation of the chiral TLL state without any intervening stripe
states \cite{wen}. Further, we neglect the possibility of the formation of
line-junction nonchiral TLLs across the vacuum regions in the shadow areas
of the metallic gates \cite{papamac}, focusing instead on the transmitted
and reflected edge states arising from the nature of the Hall fluid inside
the constriction. Thus, we devote our attention to short-ranged electronic
correlations which cause the formation of chiral Tomonaga-Luttinger liquid
(TLL) edge states (without the intervention of any stripe states
~~\cite{papastroh} arising from longer range interactions).
\par
As we will see in the following sections, such a model
of a constriction in a quantum Hall sample allows for considerable
progress to be made in developing a (quadratic) effective field theory
for the ballistic transport of current in terms of propagating chiral edge
density-wave excitations. Interesting consequences for quasiparticle
tunneling will then be shown to result from the exponentiation of these
quadratic fields, in particular, giving rise to the competition between
two RG relevant quasiparticle tunneling operators which determine the fate
of the low-bias transmission and reflection conductances through
the constriction.
In this way, we will show how our model is able to provide a
qualitative understanding of the puzzling findings of the experiments
mentioned above in a unified manner.
While it appears difficult at first to formulate
a continuum model describing a scenario of intermediate ballistic transmission
of current through such a constriction by a quadratic bosonic field theory
similar to that of Wen \cite{wen}, we find that considerable progress
can be made by understanding the role of matching (or boundary) conditions
in such a theory. In this way, we are able to set up in the following section
a very general Hamiltonian, as well as action, formalism describing transport
through such a constriction system.

\section{Continuum theory for the constriction system}
In this section, we develop a continuum theory for the model of
the constriction system presented above. However, for the sake of clarity and
continuity, we
begin by presenting the basic ingredients of Wen's continuum theory for the
infinitely long chiral-Tomonaga Luttinger liquid~~\cite{wen}.

\subsection{Continuum theory for infinite chiral TLL}
Wen's hydrodynamic
formulation describes the excitations of such a system in terms of chiral
bosonic density wave modes. The Hamiltonian (and the action) is
quadratic in the bosonic field $\phi (x,\tau)$ (where $\tau$ is the
Euclidean time) and has two parameters: the edge velocity $v$ and the filling
fraction $\nu$. This is shown below in Fig.(\ref{oneedge}).
\begin{figure}[htb]
\begin{center}
\scalebox{0.55}{
\psfrag{1}[bl][bl][2.4][0]{$\rho(x)$}
\psfrag{2}[bl][bl][2.4][0]{$j(x)$}
\psfrag{3}[bl][bl][3][0]{$x$}
\psfrag{4}[bl][bl][2.4][0]{$\phi(x,\tau)$}
\includegraphics{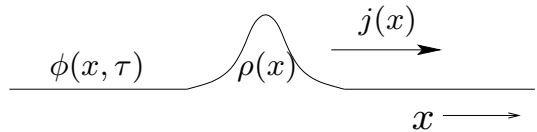}}
\end{center}
\caption{A schematic diagram for the infinitely long right-moving quantum Hall
edge state. The edge displacement is given by the bosonic field $\phi(x,\tau)$
while the edge density and current are given by $\rho(x,\tau)\sim\partial_{x}\phi$
and $j(x,\tau)\sim\partial_{\tau}\phi$ respectively.}
\label{oneedge}
\end{figure}
\par
The energy cost for density distortions of the edge of the quantum Hall system
were shown by Wen to lead to a Hamiltonian (for, say, the right-moving
edge of a Hall bar)
\beq
H = \frac{v}{4\pi\nu}\int_{-\infty}^{\infty} dx (\partial_{x}\phi_{R} (x,\tau))^{2}~.
\eeq
The equal-time (Kac-Moody) commutation relation for the bosonic field $\phi_{R}$
is given by
\beq
[\phi_{R} (x), \partial_{x}\phi_{R} (x')] = i\pi\nu\delta(x-x')~,
\eeq
which makes $\partial_{x}\phi_{R}$ the momentum canonically conjugate
to $\phi_{R}$. The edge density distortion is given by
$\rho (x) = \partial_{x}\phi_{R}(x)/(2\pi)$ and the Hamilton equation of
motion gives
\beq
i\partial_{\tau}\rho_{R}= i[H,\rho_{R}]= -v\partial_{x}\rho_{R}(x,\tau)~.
\eeq
This gives us that the density $\rho_{R} (x,\tau) = \rho_{R} (x + iv\tau)$.
Further, from the equation of continuity
\beq
i\partial_{\tau}\rho + \partial_{x} j = 0~,
\eeq
we find the current density as $j_{R} = -i\partial_{\tau}\phi_{R}/(2\pi)$. Fourier
transforming the equation of motion gives us the expected linear dispersion
relation for the edge density waves as $\omega = v k$. From the commutation
relations, we obtain the Legendre transformation for the Hamiltonian
$H(\phi_{R})$. This leads to the Euclidean action for the chiral (right moving)
TLL as
\beq
S_{R} = \frac{1}{4\pi\nu}\int_{0}^{\beta} d\tau\int_{-\infty}^{\infty} dx
\partial_{x}\phi_{R}(i\partial_{\tau} + v\partial_{x})\phi_{R} (x,\tau)~.
\eeq
The Hamiltonian for the left-moving edge density wave is the same as that
given above by for $\phi_{R}\rightarrow\phi_{L}$, but the density
$\rho_{L}= -\partial_{x}\phi_{L}/(2\pi)$. As the equal-time commutation relation
$[\phi_{L}(x), \partial_{x}\phi_{L}(x')]= -i\pi\nu\delta (x-x')$,
the action for the left moving edge chiral TLL has a Legendre transformation
term $-i\partial_{\tau}\phi_{L}\partial_{x}\phi_{L}$.

\subsection{Continuum theory for the constriction edge model}
We now formulate a continuum theory for the constriction edge model discussed
in section II along the lines of the Wen hydrodynamic description described
just above. The aim will, therefore, be to develop a quadratic theory in bosonic
fields in an edge model consisting of chiral, current carrying gapless
edge-density wave excitations describing ballistic transport through the
transmitting and reflecting edges states surrounding the constriction region.
This is shown in Fig.(\ref{beam2}) below. As discussed earlier, such a model is
critically needed in order to describe the experimentally observed scenario of
intermediate ballistic transmission through the constriction~~\cite{roddaro1}.
\begin{figure}[htb]
\begin{center}
\scalebox{0.5}{
\psfrag{1}[bl][bl][3][0]{$1,in$}
\psfrag{2}[bl][bl][3][0]{$1,out$}
\psfrag{3}[bl][bl][3][0]{$2,in$}
\psfrag{4}[bl][bl][3][0]{$2,out$}
\psfrag{5}[bl][bl][3][0]{$u$}
\psfrag{6}[bl][bl][3][0]{$r$}
\psfrag{7}[bl][bl][3][0]{$d$}
\psfrag{8}[bl][bl][3][0]{$l$}
\psfrag{9}[bl][bl][3][0]{$\nu_{2}$}
\psfrag{10}[bl][bl][3][0]{$\nu_{1}$}
\psfrag{11}[bl][bl][3][0]{$\nu_{1}$}
\psfrag{12}[c][c][2.5][0]{$(x\sim 0)$}
\includegraphics{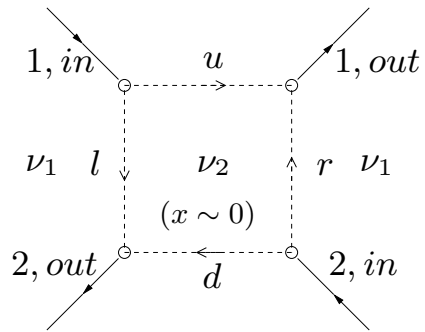}
}
\end{center}
\caption{A schematic diagram of the ``constriction" system given
by the dashed box around the region $x\sim 0$ and symbolised by the filling
fraction $\nu_{2}$ lower than that of the bulk, $\nu_{1}$. The
four chiral fields approaching and leaving this region are shown by
the arrows marked as $1,in$, $1,out$, $2,in$ and $2,out$. The dashed
horizontal and vertical lines at the junction represent the edge states
which are transmitted $(u,d)$ and reflected $(l,r)$
at the constriction respectively.}
\label{beam2}
\end{figure}
We take the spatial extent of the constriction region $2a$ to lie in
the range $l_{B}<<2a<<L$,
where $L$ is the total system size and $l_{B}$ is the magnetic length;
the external arms $(1in,\ldots,2out)$ meet the internal ones $(u,\ldots,l)$
at the four corners of the constriction. From our earlier discussions, it
is also evident that $\nu_{1}$ governs the properties of the four outer
arms while $\nu_{2}$
that of the upper and lower (transmitted) arms of the circuit at the
constriction. The effective filling factor for the right and left
(reflected) arms of the circuit $(\nu_{ref})$ is treated as a parameter
to be determined. We focus in this work on the effects of a changing filling
fraction, keeping the edge velocity $v$ the same everywhere.
\par
We will now set forth the Hamiltonian formulation of the model. This approach
will elucidate the importance of matching (or boundary) conditions in
providing a correct and consistent description of the dynamics of the system~\cite{epl}.
We will follow this up by providing the more elegant formulation of the
problem based on the action, showing how the information content of the
boundary terms is already included in this language.

\subsubsection{Hamiltonian formulation and matching conditions}
The energy cost
for chiral density-wave excitations that describe ballistic transport
in the various arms of the circuit shown in fig.(\ref{beam2}) is given by
a Hamiltonian $H = H^{ext} + H^{int}$ where
\bea
H^{ext} &=& \frac{\pi v}{\nu_{1}}[\int_{-L}^{-a}\hspace*{-0.5cm}
dx~(\rho_{1in}^{2}
+\rho_{2out}^{2}) + \int_{a}^{L}\hspace*{-0.3cm}
dx~(\rho_{2in}^{2}+\rho_{1out}^{2})]~,\nonum\\
H^{int} &=& \frac{\pi v}{\nu_{2}}\int_{-a}^{a}
dx~(\rho_{u}^{2} +\rho_{d}^{2})
+ \frac{\pi v}{\nu_{ref}}\int_{-a}^{a}\hspace*{-0.3cm}
dy~(\rho_{r}^{2}+\rho_{l}^{2})~.
\label{discham}
\eea
The densities $\rho$ are, as usual, represented in terms
of bosonic fields $\phi$ describing the edge displacement~\cite{wen}
\bea
\rho_{1in} &=& 1/2\pi \partial_{x}\phi^{1in}, \rho_{1out}= 1/2\pi
\partial_{x}\phi^{1out}\nonum\\
\rho_{2in} &=& -1/2\pi \partial_{x}\phi^{2in}, \rho_{2out} = -1/2\pi
\partial_{x}\phi^{2out}\nonum\\
\rho_{u} &=& 1/2\pi \partial_{x}\phi^{u}, \rho_{d}= -1/2\pi
\partial_{x}\phi^{d}\nonum\\
\rho_{l} &=& 1/2\pi \partial_{y}\phi^{l}, \rho_{r} = -1/2\pi \partial_{y}\phi^{r}~.
\eea
The commutation relations satisfied by these fields are familiar
\bea
[\phi^{1in}(x),\partial_{x}\phi^{1in}(x')] &=& i\pi\nu_{1}\delta(x-x')\nonum\\
&=& -[\phi^{2out}(x),\partial_{x}\phi^{2out}(x')],\nonum
\eea
\bea
[\phi^{1out}(x),\partial_{x}\phi^{1out}(x')] &=& i\pi\nu_{1}\delta(x-x')\nonum\\
&=& -[\phi^{2in}(x),\partial_{x}\phi^{2in}(x')],\nonum
\eea
\bea
[\phi^{u}(x),\partial_{x}\phi^{u}(x')] &=& i\pi\nu_{2}\delta(x-x')\nonum\\
&=& -[\phi^{d}(x),\partial_{x}\phi^{d}(x')],\nonum
\eea
\bea
[\phi^{l}(y),\partial_{y}\phi^{l}(y')] &=& i\pi\nu_{ref}\delta(y-y')\nonum\\
&=& -[\phi^{r}(y),\partial_{y}\phi^{r}(y')]~.
\label{kacmoody}
\eea
Further, the Hamiltonian equations of motion derived from $H$ again describe the
ballistic transport of chiral edge density waves
\bea
(\partial_{t} - v\partial_{x})\rho^{1in}(x,t)=&0&=(\partial_{t} -
v\partial_{x})\rho^{1out}(x,t)\nonum\\
(\partial_{t} + v\partial_{x})\rho^{2in}(x,t)=&0&=(\partial_{t} +
v\partial_{x})\rho^{2out}(x,t)\nonum\\
(\partial_{t} - v\partial_{x})\rho^{u}(x,t)=&0&=(\partial_{t} +
v\partial_{x})\rho^{d}(x,t)\nonum\\
(\partial_{t} - v\partial_{y})\rho^{l}(y,t)=&0&=(\partial_{t} +
v\partial_{y})\rho^{r}(y,t)~.
\eea
\par
The $H$ given above,
however, needs to be supplemented with matching conditions at the corners
of the constriction for a complete description. From the form
of $H$, it is clear that we need two matching conditions at
each corner; a reasonable choice is one defined on the fields and
one on their spatial derivatives. We choose, for instance, at the top-left
corner
\bea
\phi^{1in}(x=-a) &=& \phi^{u}(x=-a) + \phi^{l}(y=-a)\nonum\\
\partial_{x}\phi^{1in}(x=-a) &=& \partial_{x}\phi^{u}(x=-a)
+ \partial_{y}\phi^{l}(y=-a)
\label{match}
\eea
where $x$ and $y$ are the spatial coordinates describing the $(1in,u)$
and $l$ arms respectively. Similarly, we choose the following matching conditions
at the other three corners as
\bea
\phi^{1out}(x=a) &=& \phi^{u}(x=a) + \phi^{r}(y=-a)\nonum\\
\partial_{x}\phi^{1out}(x=a) &=& \partial_{x}\phi^{u}(x=a)
+ \partial_{y}\phi^{r}(y=-a)\nonum\\
\phi^{2in}(x=a) &=& \phi^{d}(x=a) + \phi^{r}(y=a)\nonum\\
\partial_{x}\phi^{2in}(x=a) &=& \partial_{x}\phi^{d}(x=a)
+ \partial_{y}\phi^{r}(y=a)\nonum\\
\phi^{2out}(x=-a) &=& \phi^{d}(x=-a) + \phi^{l}(y=a)\nonum\\
\partial_{x}\phi^{2out}(x=-a) &=& \partial_{x}\phi^{d}(x=-a)
+ \partial_{y}\phi^{l}(y=a)~.
\label{match2}
\eea
The equation of continuity leads to the familiar form for the current operator
$j^{\alpha}=-i\partial_{\tau}\phi^{\alpha}/(2\pi)$, where $\alpha=(1in,1out,\ldots,l,r)$.
Thus, we can easily see that current conservation at every corner arises from the
matching conditions on the bosonic fields $\phi$. While the transmitting chiral edge
modes convey a finite current across the constriction, the reflecting chiral edge modes
convey a finite ``backscattered" current across the sample. In this way, we formally
establish the intermediate ballistic transmission scenario as observed
in the experiments. Charge density fluctuations at each corner are described
by the matching conditions on $\partial_{x}\phi$. This matching condition is
a statement of the conservation of net charge density at each corner.
In this way, the two sets of matching conditions together establish the continuity
of current and charge density at every corner of the junction system.
\par
Using eqs.(\ref{match}), we compute the commutation relation
\beq
[\phi^{l},\partial_{y}\phi^{l}]_{y\rightarrow-a}
= ([\phi^{1in},\partial_{x}\phi^{1in}] -
[\phi^{u},\partial_{x}\phi^{u}])_{x\rightarrow-a}~,
\eeq
giving us $\nu_{ref}=\nu_{1}-\nu_{2}$. The commutation relation for
$\phi^{r}(y\rightarrow a)$ similarly
yields $\nu_{ref}=\nu_{1}-\nu_{2}$ once again. This is in conformity with
our result for $\nu_{ref}$ from the Landauer-Buttiker calculation.
We now demonstrate explicitly that the cases of a perfect Hall bar
($\nu_{2}=\nu_{1}$) and two Hall bubbles separated by vacuum
($\nu_{2}=0$) can be modeled as special limiting cases of the matching
conditions (eqs.(\ref{match})) given earlier. For $\nu_{1}=\nu_{2}$,
the commutation relation of the
reflecting edge states vanishes, killing its dynamics. This can also
be understood within a hydrodynamic prescription \cite{wen}, where a
vanishing effective filling factor (the amplitude of the
Kac-Moody commutation relation,
eq.(\ref{kacmoody})) leads to a diverging energy cost for edge charge
density fluctuations; the dynamics of the bosonic field characterising
such fluctuations is thus completely damped. Thus,
the reflecting edge states carry no current, while the transmitting
edge states perfectly transmit all incoming current into the outgoing
arms on the {\it opposite} side of the constriction. The matching
conditions eqs.(\ref{match}) at the four corners are then reduced to
\bea
\phi^{1,in}(x=-a) &=& \phi^{u}(x=-a)~,~\phi^{u}(a) = \phi^{1out}(x=a),\nonum\\
\phi^{2,in}(x=a)\hspace*{-0.1cm}&=&\hspace*{-0.1cm}\phi^{d}(x=a)~,
~\phi^{d}(-a) = \phi^{2out}(x=-a),\nonum\\
\partial_{x}\phi^{1,in}(-a) &=& \partial_{x}\phi^{u}(-a)~,~
\partial_{x}\phi^{u}(a) = \partial_{x}\phi^{1out}(a),\nonum\\
\partial_{x}\phi^{2,in}(a)\hspace*{-0.1cm}&=&\hspace*{-0.1cm}\partial_{x}\phi^{d}(a)~,~
\partial_{x}\phi^{d}(-a) = \partial_{x}\phi^{2out}(-a).
\label{bcqhall1}
\eea
These identifications of the fields and their spatial derivatives lead to
the continuity conditions which underpin the hydrodynamic theory of
Wen \cite{wen,kane} for the case of the two infinite chiral edges (say, upper
and lower) of a Hall bar (with filling factor $\nu_{1}$), and
eq.(\ref{kacmoody}) then reproduces the well-known
Kac-Moody commutation relation everywhere along the edges. This is shown in
Fig.(\ref{qhall1}) below.
\begin{figure}[htb]
\begin{center}
\scalebox{0.4}{
\includegraphics{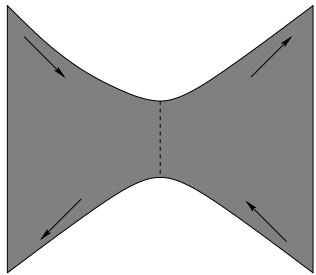}
}
\end{center}
\caption{A schematic diagram of the quantum Hall bar system with a
``constriction" which promotes quasiparticle tunneling between two
points on oppositely directed edges of the system (dashed line). The
upper and lower edges are continuous everywhere and therefore have
boundary conditions on the field $\phi$ and its spatial derivative
$\partial_{x}\phi$ as given above in equns.(\ref{bcqhall1}).}
\label{qhall1}
\end{figure}
\par
Similarly, for the case of $\nu_{2}=0$, the commutation relation for
the transmitting edge states vanishes, killing its dynamics: they carry
no current, while the reflecting edge states perfectly convey all
incoming current into the outgoing arms on the {\it same} side of the
constriction. Thus, the matching conditions eqs.(\ref{match}) at the
four corners are reduced to
\bea
\phi^{1,in}(x=-a)\hspace*{-0.1cm}&=&\hspace*{-0.1cm}\phi^{l}(y=-a)~,
~\phi^{l}(a) = \phi^{2out}(x=-a),\nonum\\
\phi^{2,in}(x=a) &=& \phi^{r}(y=a)~,~\phi^{r}(-a) = \phi^{1out}(x=a),\nonum\\
\partial_{x}\phi^{1,in}(-a) &=& \partial_{y}\phi^{l}(-a)~,~
\partial_{y}\phi^{l}(a) = \partial_{x}\phi^{2out}(-a),\nonum\\
\partial_{x}\phi^{2,in}(a) &=& \partial_{y}\phi^{r}(a)~,~
\partial_{y}\phi^{r}(-a) = \partial_{x}\phi^{1out}(a).
\label{bcqhall2}
\eea
Again, these identifications of the fields and their spatial derivatives
lead to the continuity conditions which underpin the hydrodynamic theory
of Wen \cite{wen,kane} for the case of the infinite chiral edges (say,
left and right) of two distinct Hall bubbles (each with filling factor
$\nu_{1}$) separated by vacuum, and again reproduce the familiar
Kac-Moody commutation relations everywhere along the edges. This is shown in
Fig.(\ref{qhall1}) below.
\begin{figure}[htb]
\begin{center}
\scalebox{0.5}{
\includegraphics{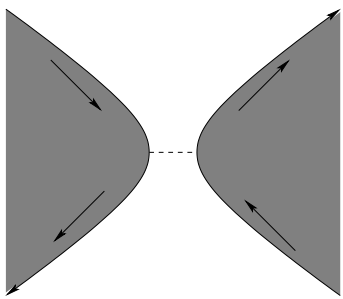}
}
\end{center}
\caption{A schematic diagram of a system of two quantum Hall droplets
separated by vacuum and with a ``constriction" which promotes electron
tunneling between two points on adjacent (and oppositely directed) edges
of the system (dashed line). The right and left edges are continuous
everywhere and therefore have boundary conditions on the field $\phi$
and its spatial derivative $\partial_{x}\phi$ as given above in
equns.(\ref{bcqhall2}).}
\label{qhall2}
\end{figure}
We have, in this way, constructed a family of free theories
describing ballistic transport through the constriction at intermediate
transmission, with those of complete transmission and reflection
representing two special cases. This represents an importance advance in generalising
the quantum impurity model of refs.\cite{kane,fendley}.

\subsubsection{Action formulation}
In this subsection, we discuss the action (or Lagrangian) formulation of our problem.
We will, in this way, demonstrate how the information content of the matching
conditions above is already encoded in the action of the system in the forms of
terms involving the local fields which are connected to one another by the matching
conditions in the Hamiltonian formalism. Thus, we begin by writing down the
action for the constriction model
\bea
S=S_{0} + S_{1} + S_{2}
\eea
where the action for the outer incoming and outgoing arms is
\bea
S_{0} &=& \int_{0}^{\beta} d\tau \int_{-\infty}^{-a} dx
\{{\cal L}_{0}[\phi_{i}^{1,in}] +
{\cal L}_{1}[\phi_{i}^{2,out}]\}\nonum\\
&&+ \int_{0}^{\beta} d\tau \int_{a}^{\infty} dx \{{\cal L}_{0}[\phi_{i}^{1,out}] +
{\cal L}_{1}[\phi_{i}^{2,in}]\}
\eea
where
\bea
{\cal L}_{0}[\phi^{\alpha}]&=&
\frac{1}{4\pi} \partial_{x}\phi^{\alpha}(i\partial_{\tau}
+ v\partial_{x})\phi^{\alpha} (x,\tau)~,\nonum\\
{\cal L}_{1}[\phi^{\alpha}]&=&
\frac{1}{4\pi} \partial_{x}\phi^{\alpha}(-i\partial_{\tau}
+ v\partial_{x})\phi^{\alpha} (x,\tau)~,
\eea
and we have normalised the entire action with regards to the bulk
filling-fraction $\nu_{1}$. Further, the action for the inner edges is
\bea
\hspace*{-0.5cm}S_{1}=\int_{0}^{\beta}\int_{-a}^{a}&[&
\frac{f}{4\pi} \partial_{x}\phi^{u}(i\partial_{\tau}
+ v\partial_{x})\phi^{u} (x,\tau)\nonum\\
&&+ \frac{f}{4\pi} \partial_{x}\phi^{d}(-i\partial_{\tau}
+ v\partial_{x})\phi^{d} (x,\tau)\nonum\\
&&+ \frac{g}{4\pi} \partial_{y}\phi^{l}(i\partial_{\tau}
+ v\partial_{y})\phi^{l} (y,\tau)\nonum\\
&&+ \frac{g}{4\pi} \partial_{y}\phi^{r}(-i\partial_{\tau}
+ v\partial_{y})\phi^{r} (y,\tau)]~,
\eea
where, by assuming that the properties of the upper and lower edge
transmitted edge states of the constriction are determined by the
effective filling-fraction
inside the constriction $\nu_{2}$, the quantity $f$ is simply given
by $f=\nu_{1}/\nu_{2}$. The quantity $g=\nu_{1}/\nu_{ref}$ (where $\nu_{ref}$
is the {\it effective} filling-fraction for the reflected edge
states on the left and right) will be determined from the analysis
presently. It is worth noting that the same information can be obtained from the
Hamiltonians (equns.(\ref{discham})) and commutation relations
(equns.(\ref{kacmoody}))together.
Finally, the action for the corner nodes is given by
\bea
&&\hspace*{-0.5cm}S_{2}=-\int_{0}^{\beta}d\tau\int_{-a}^{a}dx\int_{-a}^{a}dy\nonum\\
&&\hspace*{-0.5cm}[\delta(x+a)\delta(y+a)\partial_{x}\phi^{1in}
\{(v\partial_{x}\phi^{u}+\partial_{y}\phi^{l})
+i\partial_{\tau}(\phi^{u} + \phi^{l})\}\nonum\\
&&\hspace*{-1.0cm}+\delta(x-a)\delta(y+a)\partial_{x}\phi^{1out}
\{(v\partial_{x}\phi^{u}+\partial_{y}\phi^{r})
+i\partial_{\tau}(\phi^{u} + \phi^{r})\}\nonum\\
&&\hspace*{-1.0cm}+\delta(x-a)\delta(y-a)\partial_{x}\phi^{2in}
\{(v\partial_{x}\phi^{d}+\partial_{y}\phi^{r})
-i\partial_{\tau}(\phi^{d} + \phi^{r})\}\nonum\\
&&\hspace*{-1.0cm}+\delta(x+a)\delta(y-a)\partial_{x}\phi^{2out}
\{(v\partial_{x}\phi^{d}+\partial_{y}\phi^{l})
-i\partial_{\tau}(\phi^{d} + \phi^{l})\}]~.
\label{boundaryterms}
\eea
\par
We can now see the effects of these local terms in the action by computing the
equations of motion for the various fields
from the action. For the sake of brevity, we carry out this exercise at only the
upper-left corner. The results obtained from the other three corners are precisely
the same. Thus, we first compute the equation of motion of the ``outer" field
$\phi_{1in}(x=-a)$ by extremising the action $S$ with regards to
$\partial_{x}\phi^{1in}(x=-a)$
\bea
\frac{\delta S}{\delta(\partial_{x}\phi^{1in}_{-a})} &=&
v(\partial_{x}\phi^{1in}-\partial_{x}\phi^{u}-\partial_{y}\phi^{l})\nonum\\
&&+ i\partial_{\tau}(\phi^{1in}-\phi^{u}-\phi^{l}) = 0
\eea
where we have suppressed the dependences of the fields on the spatial
coordinates for the sake of compactness. From this, we can immediately see the
matching conditions on $\phi$ and $\partial_{x}\phi$ at $(x=-a,y=-a)$
given earlier.
We now compute the other two equations of motion at the top left corner in the
same way. We find, thus,
\bea
\frac{\delta S}{\delta(\partial_{x}\phi^{u}_{-a})} &=&
v(f\partial_{x}\phi^{u}-\partial_{x}\phi^{1in})\nonum\\
&&+ i\partial_{\tau}(f\phi^{u}-\phi^{1in}) = 0\nonum\\
\frac{\delta S}{\delta(\partial_{y}\phi^{l}_{-a})} &=&
v(g\partial_{y}\phi^{l}-\partial_{x}\phi^{1in})\nonum\\
&&+ i\partial_{\tau}(g\phi^{l}-\phi^{1in}) = 0~,
\eea
from which we can see that the currents $j^{u}(x=-a)$ and $j^{l}(y=-a)$ are
given by
\bea
j^{u}(x=-a) &=& -i\partial_{\tau}(\phi^{u} + \frac{\phi^{1in}}{f})\nonum\\
&=& v (\phi^{u} - \frac{\phi^{1in}}{f})
\equiv v\rho^{u}(x=-a)\nonum\\
j^{l}(y=-a) &=& i\partial_{\tau}(\phi^{l} + \frac{\phi^{1in}}{g})\nonum\\
&=& v (\phi^{l} - \frac{\phi^{1in}}{g})
\equiv v\rho^{l}(y=-a)~.
\eea
In the above relations, the currents $(j^{u}, j^{l})$ and
corresponding densities $(\rho^{u}, \rho^{l})$ are
those propagated from the incoming arm $1in$ into the u(pper)
and l(eft) edge states respectively. Now, by applying Kirchoff's law
for the conservation of current (or, more generally, the equation of
continuity) at the upper left corner junction,
$j^{1in}(x=-a)=j^{u}(x=-a)+j^{l}(y=-a)$, we obtain
\beq
\frac{1}{f} + \frac{1}{g} = 1~,
\eeq
which for $f=\nu_{1}/\nu_{2}$ gives $g=\nu_{1}/(\nu_{1}-\nu_{2})$. This,
then, gives us the effective filling-fraction of the reflected edge
states as $\nu_{ref}=\nu_{1}-\nu_{2}$. In this way, we can see that the
action $S$ contains all the information content given by the Hamiltonians
together with the matching conditions.

\section{Boundary theory for the constriction system}
In this section, we evaluate the role played by local inter-edge quasiparticle
tunneling processes deep inside the constriction region in determining the
fate of transport through the constriction. In order to do so, we proceed
by first integrating out all bosonic degrees of freedom except the few involved in
the tunneling processes. In this way, we are left with an effective
{\it boundary} theory~~\cite{kane}. Given that we have a Gaussian action in terms
of the bosonic fields, integrating out various bosonic degrees of freedom can
be easily accomplished by performing Gaussian integrations~~\cite{gogolin,giamarchi}.
(another analogous method involves using the solutions to the equations
of motion~~\cite{kane,lalcontacts}). As this is a very standard procedure, we
refer the reader to Refs.(\cite{gogolin,giamarchi}) for details. We pass
instead to presenting the various boundary theories obtained in our model, revealing
in turn the two interedge quasiparticle tunneling processes which compete in
determining the low energy dynamics of the system.
\par
Now, as long as there is no quantum coherence across the constriction
region, it is easily seen that the problem of weak, local quasiparticle tunneling
between the upper ({\it u}) and lower ({it d}) edges deep inside the
constriction region (at, say, $x=0$) is exactly the same as that of local
quasiparticle tunneling between the oppositely directed edges of a
homogeneous quantum Hall bar with filling fraction $\nu_{2}$~~\cite{kane}.
Importantly, the charge and statistics of the quasiparticles undergoing
such tunneling processes should be governed by the local filling fraction $\nu_{2}$ alone.
Thus, in the action formalism presented earlier, such a quasiparticle tunneling
process can be added to the action $S$ by the term
$\lambda_{1}\cos(\sqrt{\nu_{1}}(\phi^{u}(0)-\phi^{d}(0)))\equiv
\lambda_{1}\cos(\phi^{ud}(0))$, where the tunnel coupling strength is given by
$\lambda_{1}$. Integrating out all bosonic degrees of freedom but
$\phi^{ud}(x=0)$, we obtain the familiar Kane-Fisher type boundary
theory~~\cite{kane}
\beq
S_{ud} = \sum_{\on} \frac{|\on|}{2\pi\nu_{2}}|\phi^{ud}_{\on}(x=0)|^{2}
+ \hspace*{-0.1cm}\int\hspace*{-0.1cm} d\tau \lambda_{1}\cos(\phi^{ud}(x=0,\tau)).
\eeq
Applying a standard RG procedure, we find the RG equation for $\lambda_{1}$ as
\beq
\frac{d\lambda_{1}}{dl} = (1 - \nu_{2}) \lambda_{1}~.
\eeq
As $\nu_{2} < 1$, the coupling $\lambda_{1}$ is found to be RG relevant
and will grow under the flow to low energies/long lengthscales. Further,
this quasiparticle tunneling process will clearly lower the transmission
conductance across the constriction $g_{1in,1out}$ (for a source-drain
bias as shown in Fig.(2)).
\par
We have, however, at least one other local quasiparticle tunneling process to
account for: it is that between the left ({\it l}) and
right ({\it r}) edges of the constriction and is revealed by the generalised
quasiparticle-quasihole symmetry of the ground-state in the lowest
Landau level (LLL) \cite{girvin,epl}. This symmetry dictates that all
properties of a quantum Hall system composed of quasiparticles in a partially
filled lowest Landau level and with a filling factor $\nu_{qp}$ can be
equivalently described by those of a quantum Hall system composed of
quasiholes and with a filling factor $\nu_{qh}=1-\nu_{qp}$. This simple
relation between $\nu_{qp}$ and $\nu_{qh}$ can be derived easily for the
case of the filling factor (and, hence, electronic density) of the quantum
Hall system deviating from a filling factor of $\nu_{0}=1/q$
(where $q=2n+1\in{\cal Z}$)~~\cite{yoshioka}. To see this, first note that by
increasing the electronic density of the system, we add $q$ quasiparticles
for each electron added. Then, for $n_{0}$ being the original electronic
density, $n_{e}$ the new increased electronic density and $n_{qp}$ the density of
quasiparticles,
\beq
n_{e} = n_{0} + \frac{n_{qp}}{q} = \frac{\nu_{0}}{2\pi l_{B}^{2}}
+ \frac{\nu_{0}\nu_{qp}}{2\pi l_{B}^{2}}~,
\eeq
where $l_{B}$ is the magnetic length and we have used $n_{0}=\nu_{0}/(2\pi l_{B}^{2})$.
This gives us the quasiparticle filling factor $\nu_{qp}$ as
\beq
\nu_{qp} = \frac{\nu_{e}}{\nu_{0}}~,
\eeq
where $\nu_{e}=2\pi l_{B}^{2} n_{e}$ is the electronic filling factor. A
similar calculation
for the case of an equally lowered value of the electronic density $n_{e}$ can also be
carried out. We must now
remember that we add $q$ quasiholes to the system for every electron removed.
Then, by following
the same line of arguments, we get the quasihole filling factor $\nu_{qh}$ as
\beq
\nu_{qh} = 1 - \frac{\nu_{e}}{\nu_{0}} = 1 - \nu_{qp}~.
\eeq
Thus, we can see that $\nu_{qp}$ and $\nu_{qh}$ are related by a quasiparticle-quasihole
conjugation transformation. Further, for the case of $\nu_{0}=1$, $\nu_{qp}=\nu_{e}$
and $\nu_{qh}=\nu_{h}=1-\nu_{e}$ are the well-known electron and hole conjugation
symmetric filling factors with respect to the completely filled lowest electronic Landau
level~~\cite{girvin}.
\par
\begin{figure}[htb]
\begin{center}
\scalebox{0.3}{
\psfrag{1}[bl][bl][5][0]{$\frac{\nu_{2}}{\nu_{1}}$}
\psfrag{2}[bl][bl][4][0]{$0$}
\psfrag{3}[bl][bl][4][0]{$0$}
\psfrag{4}[bl][bl][4][0]{$1$}
\psfrag{5}[bl][bl][4][0]{$1$}
\psfrag{6}[bl][bl][4][0]{(a)}
\psfrag{7}[bl][bl][4.9][0]{$1-\frac{\nu_{2}}{\nu_{1}}$}
\psfrag{8}[bl][bl][4][0]{$1$}
\psfrag{9}[bl][bl][4][0]{$1$}
\psfrag{10}[bl][bl][4][0]{$0$}
\psfrag{11}[bl][bl][4][0]{$0$}
\psfrag{12}[c][c][4][0]{(b)}
\psfrag{13}[c][c][4][0]{(c)}
\psfrag{14}[c][c][4][0]{(d)}
\includegraphics{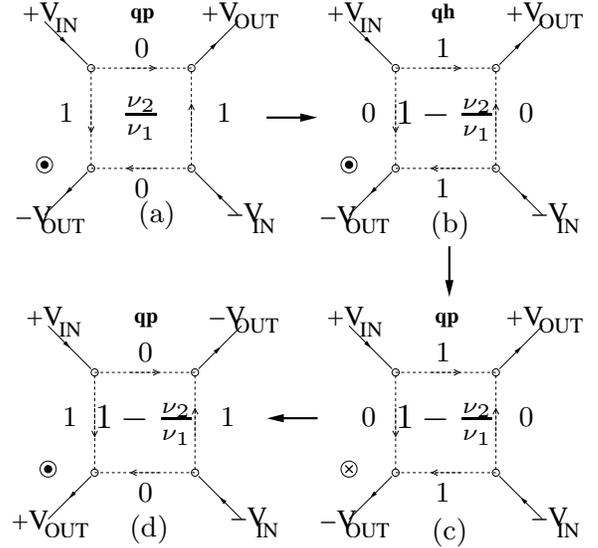}
}
\end{center}
\caption{The quasiparticle-quasihole (qp-qh) symmetry of the
$(\nu_{1},\nu_{2})$ constriction geometry in terms of the relative filling
factor. The source-drain bias $2V_{in}$ is applied to the
two incoming arms while $\pm V_{out}$ are the equilibration potentials
of the two outgoing arms. See text for a detailed description of the
transformations linking (a) to (d).}
\label{qpqh}
\end{figure}

The application of
this symmetry to the constriction model with a spatially dependent filling
fraction relies on (a) the fact that the
fractional (integer) quantum Hall ground state in the bulk of the system can be
thought of as the completely filled effective lowest Landau level of
quasiparticles (electrons), (b) that this state is protected by an energy gap which
is larger than all other energy scales in the problem and (c) there is no
Landau level mixing.
While the argument for a constriction circuit with the bulk being in the
integer quantum Hall state of $\nu_{1}=1$ has been given in
Ref.(\cite{roddaro1}), a generalisation for any constriction circuit with a
general bulk filling factor $\nu_{1}<1$ was presented in Ref.(\cite{epl}).
The argument is recounted below and encapsulated in Fig.(\ref{qpqh}).
\par
By scaling the filling factors of all
regions by $\nu_{1}$, we now have the effective filling factor
of the bulk as $1$ and that of the quantum Hall ground
state inside the constriction region as the relative filling fraction
$f^{-1}=\nu_{2}/\nu_{1}$. Carrying out the qparticle-qhole conjugation
transformation, we go to a system of holes (with the direction of the
external magnetic field unaffected) but with the relative filling
fraction of the constriction
now given by $g^{-1}=1-f^{-1}$. We can then map this system of qholes
onto that of
time-reversed qparticles (i.e., qparticles in an oppositely directed external
magnetic field). Finally, by rotating the system by $180^{o}$ around
the axis of the two outgoing current directions, we are left with a system
of electrons with the external magnetic field pointing in the original
direction. This final system is, however, crucially different in two ways
from the original one. First, the filling factor of the
fractional quantum Hall ground state inside the constriction region
has, as noted earlier, changed from $f^{-1}$ to $g^{-1}=1-f^{-1}$. Second,
the directions transmitted and reflected currents has been interchanged in
going from the original system to the final one. It is easy to show that
current conservation dictates that \cite{roddaro1}
\beq
I_{f^{-1}} + I_{g^{-1}} = 2\nu_{1}\frac{e^{2}}{h} V_{in}~,
\eeq
where $I_{f^{-1}}$ and $I_{g^{-1}}$ are the transmitted currents in the
original system and the final system after the mappings (i.e., the
reflected current in the original system). Relations can also be written
down in terms of differential transmission $t=dI/dV$ and reflection $r$
(in units of $\nu_{1}e^{2}/h$)
\bea
t_{f^{-1}} &=& 1 - t_{g^{-1}} = r_{g^{-1}}\nonum\\
t_{g^{-1}} &=& 1 - t_{f^{-1}} = r_{f^{-1}}~.
\eea
This argument also clearly demonstrates that the local tunneling process
that transports these qholes between the left ({\it l}) and right ({\it r})
edges of the constriction system is governed by the effective filling-fraction
of $g^{-1} = 1 - \nu_{2}/\nu_{1}$. Thus, we denote such a local tunneling process
(again, chosen at $y=0$) on the {\it l} and {\it r}
edges by a weak tunnel coupling $\lambda_{2}$ and a term in the action $S$,
$\lambda_{2}\cos(\phi^{l}(0) - \phi^{r}(0))\equiv\lambda_{2}\cos(\phi^{lr}(0))$.
Integrating out all bosonic degrees of freedom but $\phi^{lr}(y=0)$, we
obtain the another Kane-Fisher type boundary theory~~\cite{kane}
\beq
S_{lr} = \sum_{\on} \frac{|\on|}{2\pi g}|\phi^{lr}_{\on}(y=0)|^{2}
+ \int d\tau \lambda_{2}\cos(\phi^{lr}(y=0,\tau))~.
\eeq
Thus, we can see that the RG equation for the coupling $\lambda_{2}$
is given by
\bea
\frac{d\lambda_{2}}{dl} &=& (1-\frac{1}{g})\lambda_{2}\nonum\\
&=& (1 - (1-\frac{\nu_{2}}{\nu_{1}}))\lambda_{2}
= \frac{\nu_{2}}{\nu_{1}}\lambda_{2}~.
\eea
As both $(\nu_{1},\nu_{2})>1$ and $\nu_{2}<\nu_{1}$, we can see that the
coupling $\lambda_{2}$ is also RG relevant and will grow under the flow
to low energies/long lengthscales. Further, this tunneling process will
increase the transmission conductance $g_{1in,1out}$.
\par
Since we have two RG relevant boundary operator couplings which
affect the transmission conductance across the constriction in opposite
ways, we need to determine the conditions under which one wins over the
other. From the scaling dimensions of the two operators (as employed
in their respective RG equations), we can see that the two couplings
grow equally fast for a critical $\nu^{*}_{2}$
\beq
\nu^{*}_{2} = \frac{\nu_{1}}{1+\nu_{1}}~.
\eeq
For this critical $\nu^{*}_{2}$, then, the transmission and reflection
conductances will be held fixed by the generalised qparticle-qhole symmetry
all along the RG flow from weak to strong coupling! For $\nu_{2} < \nu^{*}_{2}$,
$\lambda_{1}$ dominates over $\lambda_{2}$, which will lead to a minimum
of the transmission conductance (i.e., a maximum in the reflection
conductance) at low energies (bias/temperature) given by the bulk
conductance $\nu_{1}$. The quantum Hall constriction system will then resemble that
of two quantum Hall droplets separated by vacuum, shown in Fig.(\ref{qhall2}) above.
Similarly, for $\nu_{2} > \nu^{*}_{2}$,
$\lambda_{2}$ dominates over $\lambda_{1}$ and will lead to the opposite
case of a maximum in the transmission conductance (i.e., a minimum in the
reflection conductance) at low energies (bias/temperature) given by the
bulk conductance $\nu_{1}$. The quantum Hall constriction system will then resemble that
of a single quantum Hall bar, shown in Fig.(\ref{qhall1}) above. As was discussed
in detail in an earlier section, these were indeed many of the puzzling experimental
findings of Refs. \cite{roddaro1,roddaro2,chung}.
\par
We can see that, for the critical value of $\nu^{*}_{2}$ predicted by
our theory, the symmetry-determined (i.e., energy scale independent)
constriction transmission conductance is given by
\bea
t(\nu^{*}_{2})&=&\frac{g_{1in,1out}(\nu^{*}_{2})}{G_{b}}
=\frac{\nu^{*}_{2}}{\nu_{1}} = \frac{1}{1+\nu_{1}}\nonum\\
&=& 1-\nu_{2}^{*}~.
\eea
We now present the above results for the first three generations of the
heirarchical sequence of quantum Hall states.\\
(i) For the bulk filling-factor belonging to the primary sequence
$\nu_{1}=1/(2p-1)$, we obtain
\beq
\nu^{*}_{2}=\frac{1}{2p}~~,~~t(\nu^{*}_{2})=\frac{2p-1}{2p}~.
\eeq
Specifically, for
$\nu_{1}=1$, we get $\nu^{*}_{2} = 1/2 = t(\nu^{*}_{2})$ and for
$\nu_{1}=1/3$, we get $\nu^{*}_{2} = 1/4$, $t(\nu^{*}_{2}) = 3/4$. Both
these sets of results match the experimental findings of Refs.
\cite{roddaro1,roddaro2,chung}.\\
(ii)Further, we find that for the case of
the second generation of the heirarchical states, $\nu_{1}=2p/(2pq\pm 1)$
(where $p=1$,$q=3,5,\ldots$), we find
\beq
\nu_{2}^{*} = \frac{2p}{2p(1+q)\pm 1}~~,~~
t(\nu_{2}^{*})=\frac{2pq\pm 1}{2p(1+q)\pm 1}~.
\eeq
Specifically, for the case
of $\nu_{1}=2/5$, $\nu_{2}^{*}=2/7$ and $t(\nu_{2}^{*})=5/7$.\\
(iii) Extending
these results to the third generation of the heirarchical states,
$\nu_{1}=4p_{1}p_{2}/(q(4p_{1}p_{2}\pm 1)\pm2p_{2})$
(where $p_{1}=1=p_{2}$,$q=3,5,\ldots$), we obtain
\bea
\nu_{2}^{*}&=&\frac{4p_{1}p_{2}\pm 1}{((1+q)(4p_{1}p_{2}\pm 1) \pm 2p_{2})}~,\nonum\\
t(\nu_{2}^{*})&=&\frac{q(4p_{1}p_{2}\pm 1)\pm 2p_{2}}
{(1+q)(4p_{1}p_{2}\pm 1) \pm 2p_{2}}~.
\eea
Specifically, for the case of
$\nu_{1}=3/7$, $\nu_{2}^{*}=3/10$ and $t(\nu_{2}^{*})=7/10$.
The results presented for the cases of $\nu_{1}=2/5, 3/7$ can be tested experimentally,
and we will comment on this in a later section.
\begin{figure}[htb]
\begin{center}
\scalebox{0.4}{
\psfrag{1}[bl][bl][4][0]{$\ln\lambda_{1}$}
\psfrag{2}[bl][bl][4][0]{$\ln\lambda_{2}$}
\psfrag{3}[bl][bl][4][0]{$\nu_{2}<\nu_{2}^{*}$}
\psfrag{4}[bl][bl][4][0]{$\nu_{2}>\nu_{2}^{*}$}
\psfrag{5}[bl][bl][4][0]{$\nu_{2}=\nu_{2}^{*}$}
\includegraphics{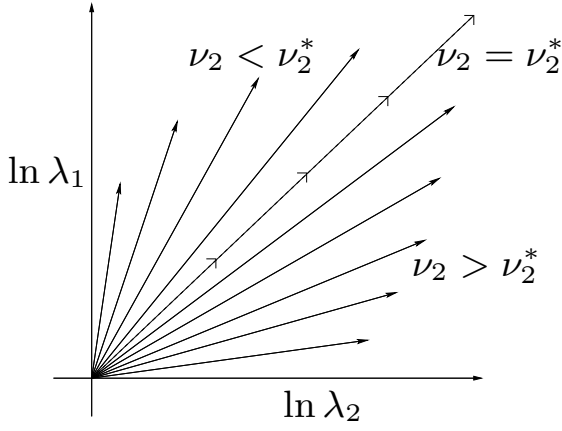}
}
\end{center}
\caption{The RG phase diagram for the model as a plot of the
function $\ln\lambda_{1}/\ln\lambda_{2} = \nu_{1}(1/\nu_{2} - 1)$.
All RG flows lead away from the weak-coupling unstable fixed point at
the origin. Properties of the dashed critical line and regions above
$(\nu_{2}<\nu_{2}*)$ and below $(\nu_{2}>\nu_{2}^{*})$ are explained
in the text.}
\label{rgph}
\end{figure}
\par
Finally, we present the RG phase diagram of the model in Fig.(\ref{rgph})
given above. The origin represents the family
of weak-coupling fixed point theories at partial transmission described
earlier, while the RG flows are to the familiar fixed point theories
\cite{kane,fendley} of complete reflection ($\nu_{2}<\nu_{2}^{*}$, see Fig.(\ref{qhall2})
given above), complete transmission ($\nu_{2}>\nu_{2}^{*}$, see Fig.(\ref{qhall1})
given above) and to a new symmetry dictated fixed point theory on the diagonal
($\nu_{2}=\nu_{2}^{*}$). The diagonal is, in fact, a separatrix -- a line of gapless
critical theories all possessing the quasiparticle-quasihole symmetry described above --
dividing RG flows to a metallic phase (as evidenced by the perfect transmission through
the constriction) and an insulating phase (as seen by the perfect reflection at the
constriction) at strong coupling. The qparticle-qhole symmetry of the constriction system
can also be seen in the reflection symmetry of the RG flows in the two segments on either
side of the separatrix: in physical terms, this means that while the upper (lower)
segment
represent RG flows towards a qparticle insulator (metal), the picture is exactly reversed
for a description in terms of qholes. This is also amply clear in terms the comparison of
the two scaling dimensions: this analysis answers the question as to which of
the qparticle
and qhole boundary degrees of freedom (given above in the two boundary theories for
qparticle and qhole tunneling respectively) becomes massive first, thereby
allowing the remaining gapless boundary degrees of freedom to determine
the low energy, long wavelength dynamics of the quantum Hall constriction system at
strong coupling. Finally, this symmetry of the RG phase diagram is reminiscent of
the edge-state transmission duality~\cite{shimshoni,xiong} that is known to exist
in the Chalker-Coddington model~\cite{chalcodd} as applied to the study of the
quantum Hall transitions.
\par
The novel structure of the RG phase diagram, thus,
reflects on the fact that the experimentally observed gate-voltage tuned
metal-insulator transition (at vanishing edge bias) is in fact
shaped not only by boundary critical phenomena (i.e., local interedge quasiparticle
tunneling processes, relying on an interplay of the physics of impurity scattering and
electron electron interactions~~\cite{kane}), but also by the presence of a global
symmetry (i.e., the quasiparticle-quasihole conjugation symmetry) and the requirement
of overall current conservation in the system.
These findings highlight the novel generalisation of the quantum impurity problem
of refs.\cite{kane,fendley} that has been accomplished here. We can now proceed to a
study of the correlators and conductances of the system in the next section, with an
effort towards reinforcing the physical picture presented by these RG flows.

\section{Correlators and conductances of the constriction model}
In this section, we present computations of various density-density correlators
of the fields in the
constriction model for the three cases of weak coupling ballistic transport (i.e., no
interedge tunneling), strong coupling with interedge tunneling for $\nu_{2}<\nu_{2}^{*}$
and strong coupling with interedge tunneling for $\nu_{2}>\nu_{2}^{*}$. We then employ
these correlators in a Kubo formulation to compute the chiral linear dc conductances
of the
system. In this way, we will confirm the physical picture of the dip-to-peak evolution
developed in the last section. We will, in this way, also be able to see the consequence
of the qparticle-qhole symmetry on measurable quantities like conductances, confirming
the physical picture of transport through the constriction system presented earlier.
\par
In all that follows, we switch from the Euclidean time $\tau$ to Matsubara
frequencies $\on$.
This will also be seen to facilitate the computation of the the linear dc conductances.
Thus, we begin by computing certain density-density correlators, e.g.,
$\langle[\partial_{x}\phi^{1in}_{\on}(x),
\partial_{x}\phi^{1out}_{-\on}(x')]\rangle$, for
the free theory
$S$ given earlier (i.e., $S$ in the absence of all interedge tunneling processes)
\bea
&&\hspace*{-1cm}\langle[\partial_{x}\phi^{1in}_{\on}(x),
\partial_{x}\phi^{1out}_{-\on}(x')]\rangle =\nonum\\
&&\hspace*{-1cm}\langle[(\partial_{x}\phi^{u}_{-a}+\partial_{y}\phi^{l}_{-a})
(\partial_{x}\phi^{u}_{a}
+\partial_{y}\phi^{r}_{-a})]\rangle e^{-|\on(x'-x-2a)/v|}\nonum\\
&&\hspace*{-1cm}=\langle[\partial_{x}\phi^{u}_{-a},\partial_{x}\phi^{u}_{-a}]
\rangle e^{-|\on(x'-x)/v|}\nonum\\
&&\hspace*{-1cm}=\frac{2\pi\nu_{2}}{v^{2}}|\on|e^{-|\on(x'-x)/v|}~,
\eea
where we have used the commutation relations for the various fields and the
fact that all transport on the various edges is ballistic and described
by the solutions to the chiral equations of motion for the edge density waves
given earlier.
Further, for the sake of notational brevity, we suppressed the $\on$ frequencies
in all subscripts on the right hand side, keeping
only the
spatial dependence in the subscripts in the correlator expressions.
The $e^{-|\on(x'-x)/v|}$ factor is the expected phase (easily seen upon
performing an analytic continuation to real frequencies $\omega$) associated
with the ballistic
transport between the points $x$ and $x'$. As we will soon see when deriving
the expressions for
the linear dc conductances, this phase factor vanishes upon taking the limit
of vanishing
frequencies, while the filling factor dependence is crucial.
\par
In the same way, we find the other density-density correlators between
the fields outside the constriction as
\bea
&&\hspace*{-1.0cm}\langle[\partial_{x}\phi^{1in}_{\on}(x),
\partial_{x}\phi^{2out}_{-\on}(x')]\rangle =
-\frac{2\pi(\nu_{1}-\nu_{2})}{v^{2}}|\on|e^{-|\on(x'-x)/v|},\nonum\\
&&\hspace*{-0.7cm}\langle[\partial_{x}\phi^{2in}_{\on}(x),
\partial_{x}\phi^{2out}_{-\on}(x')]\rangle =
-\frac{2\pi\nu_{2}}{v^{2}}|\on|e^{-|\on(x'-x)/v|}~,\nonum\\
&&\hspace*{-0.7cm}\langle[\partial_{x}\phi^{2in}_{\on}(x),
\partial_{x}\phi^{1out}_{-\on}(x')]\rangle =
\frac{2\pi(\nu_{1}-\nu_{2})}{v^{2}}|\on|e^{-|\on(x'-x)/v|}.
\eea
We now turn to the case of strong coupling interedge tunneling within the
constriction. Here,
three scenarios can be realised, and we study each of them in turn. First,
for the case of
$\nu_{2}<\nu_{2}^{*}$, we have already seen that quasiparticle tunneling between the
upper and lower edges of the constriction region dominates at strong coupling.
Then, from
the boundary theory given earlier, it is clear that this strong coupling scenario
possesses another boundary condition (dynamically generated due to the RG flow):
$\phi^{u}(x=0) = \phi^{d}(x=0)$. Using this boundary
condition while computing the four density-density correlators given above, we find
\bea
&&\hspace*{-0.7cm}\langle[\partial_{x}\phi^{1in}_{\on}(x),
\partial_{x}\phi^{1out}_{-\on}(x')]\rangle =
0~,\nonum\\
&&\hspace*{-0.7cm}\langle[\partial_{x}\phi^{1in}_{\on}(x),
\partial_{x}\phi^{2out}_{-\on}(x')]\rangle =
-\frac{2\pi\nu_{1}}{v^{2}}|\on|e^{-|\on(x'-x)/v|}~,\nonum\\
&&\hspace*{-0.7cm}\langle[\partial_{x}\phi^{2in}_{\on}(x),
\partial_{x}\phi^{2out}_{-\on}(x')]\rangle =
0~,\nonum\\
&&\hspace*{-0.7cm}\langle[\partial_{x}\phi^{2in}_{\on}(x),
\partial_{x}\phi^{1out}_{-\on}(x')]\rangle =
\frac{2\pi\nu_{1}}{v^{2}}|\on|e^{-|\on(x'-x)/v|}~.
\eea
It is easy to see from these correlators that the physical system here is
that visualised in Fig.(\ref{qhall2}).
\par
Next, for the case of
$\nu_{2}>\nu_{2}^{*}$, we have already determined that quasiparticle tunneling
between the
left and right edges of the constriction region dominates at strong coupling.
Then, from
the boundary theory given earlier, it is clear that this strong coupling scenario
possesses another boundary condition (again, dynamically generated due to the RG flow):
$\phi^{l}(y=0) = \phi^{r}(y=0)$. Using this boundary
condition while computing the four density-density correlators given above, we find
\bea
&&\hspace*{-0.7cm}\langle[\partial_{x}\phi^{1in}_{\on}(x),
\partial_{x}\phi^{1out}_{-\on}(x')]\rangle =
\frac{2\pi\nu_{1}}{v^{2}}|\on|e^{-|\on(x'-x)/v|},\nonum\\
&&\hspace*{-0.7cm}\langle[\partial_{x}\phi^{1in}_{\on}(x),
\partial_{x}\phi^{2out}_{-\on}(x')]\rangle =
0~,\nonum\\
&&\hspace*{-0.7cm}\langle[\partial_{x}\phi^{2in}_{\on}(x),
\partial_{x}\phi^{2out}_{-\on}(x')]\rangle =
-\frac{2\pi\nu_{1}}{v^{2}}|\on|e^{-|\on(x'-x)/v|}~,\nonum\\
&&\hspace*{-0.7cm}\langle[\partial_{x}\phi^{2in}_{\on}(x),
\partial_{x}\phi^{1out}_{-\on}(x')]\rangle = 0~.
\eea
It is again easy to see from these correlators that the physical system here
is that visualised in
Fig.(\ref{qhall1}).
\par
Finally, we turn to considering the qparticle-qhole symmetric case of
$\nu_{2}=\nu_{2}^{*}$. It is
clear that since the two RG flows to strong coupling indicate opposing tendencies
on the system,
i.e., very different physical configurations for the system (Figs.(\ref{qhall1})
and (\ref{qhall2})
respectively), the equally fast growth of both couplings still cannot lead
to the generation of
any new boundary conditions in this case. It is then easily concluded that
all the density-density correlators
studied above must appear to be exactly the same at strong coupling as found
to be at weak coupling
(and, indeed, all along the RG flow). Thus, the consequence of the global
symmetry of qparticle-qhole
conjugation is to keep the various transmission and reflection edge fields
at the constriction
from becoming massive locally. This confirms the existence of a line of
critical (gapless) theories,
all possessing the symmetry mentioned above and unstable to relevant RG
perturbations upon
changing the parameter $\nu_{2}$ from its critical value; the correlators
confirm that the RG flow
is towards strong coupling theories where either the transmitting or the
reflecting edges become
massive locally, suppressing some of the correlators while giving the other
correlators the
values they would have in the two scenarios in Figs.(\ref{qhall1}) and (\ref{qhall2}).
\par
As we will now see, by using the fact that the charge and current densities
for a chiral edge bosonic field $\phi$ are simply related to another
another, these density-density correlators can be employed in computing several
2-terminal chiral
linear (dc) conductances. These conductances can be derived from a linear response
type Kubo
formulation~~\cite{kane,moon,kanefishreview}, yielding relations linking them
to the correlators computed above in the form of retarded response functions
(obtained upon performing an analytic continuation
from Matsubara frequencies $\on$ to real frequencies $\omega$)
\bea
g_{\alpha\beta} (x,x')&=&
\lim_{\omega\rightarrow 0}
(-1)^{[\tilde{\alpha}+\tilde{\beta}]}\frac{e^{2}v^{2}}{2\pi h\omega}\times\nonum\\
&&\hspace*{0.2cm}\times\langle[\partial_{x}\phi^{\alpha}_{\omega}(x),
\partial_{x}\phi^{\beta}_{-\omega}(x')]\rangle~,
\eea
where $(\alpha,\beta) = (1in,\ldots,2out)$ are the terminal indices,
$(\tilde{\alpha},\tilde{\beta})$ are
the terminal numbers (i.e., $1$ and $2$) associated with these terminal indices
and $[\tilde{\alpha}+\tilde{\beta}]$ is a number modulo $2$ such that the factor
$(-1)^{[\tilde{\alpha}+\tilde{\beta}]}$ restores
the direction of net current flow from source to drain as positive.
From the expressions for
the correlators given earlier, it is clear that in the dc limit
$\omega\rightarrow 0$ , the linear
conductances no longer depend on the spatial coordinates of $x$ and $x'$; this
is a consequence
of the fact that transport along the edges is ballistic and equilibriation takes
place only in
the reservoirs~~\cite{kanefishreview}. We can now simply use the various
correlators computed
above in calculating the various 2-terminal chiral linear conductances of the system.
For the sake of brevity, we summarise in table (\ref{condtable})
presented below, our calculations of the chiral linear conductances $g_{1in,1out}$
and $g_{1in,2out}$,
representing the transmission and reflection through the constriction, at weak coupling
(ballistic transport only) and the three strong coupling scenarios of
$\nu_{2}<\nu_{2}^{*}$,
$\nu_{2}>\nu_{2}^{*}$ and $\nu_{2}=\nu_{2}^{*}$.
\begin{table}[htb]
\begin{tabular}{p{2.7cm} c c}
\hline
Constriction filling \hspace*{1.0cm}$\nu_{2}$&\hspace*{0.2cm} $g_{1in,1out}
(e^{2}/h)$ &\hspace*{0.2cm} $g_{1in,2out} (e^{2}/h)$\\
\hline
Weak coupling & $\nu_{2}$ & $\nu_{1}-\nu_{2}$\\
Strong coupling\\$(\nu_{2}<\nu_{2}^{*})$ & 0 & $\nu_{1}$\\
Strong coupling\\$(\nu_{2}>\nu_{2}^{*})$ & $\nu_{1}$ & 0\\
Strong coupling\\$(\nu_{2}=\nu_{2}^{*})$ & $\nu_{2}$ & $\nu_{1}-\nu_{2}$\\
\hline
\end{tabular}
\caption{Values of two chiral linear conductances, $g_{1in,1out}$ and $g_{1in,2out}$,
representing the transmission and reflection through the constriction at weak coupling
(ballistic transport only) and the three strong coupling scenarios of
$\nu_{2}<\nu_{2}^{*}$,
$\nu_{2}>\nu_{2}^{*}$ and $\nu_{2}=\nu_{2}^{*}$.}
\label{condtable}
\end{table}
\par
The other two conductances $g_{2in,1out}$
and $g_{2in,2out}$ can be computed in precisely the same manner. The physical
picture of weak
coupling ballistic transport and the three strong coupling scenarios presented
earlier is
immediately confirmed from the expressions given in table (\ref{condtable}).
The finite temperature $T$ (or voltage $V$) expressions for the perturbative
corrections to the weak coupling
ballistic chiral conductances due to the qparticle and qhole interedge
tunneling processes
revealed earlier can also be computed from the boundary theories presented
earlier~~\cite{kane}. Thus, we find that
\bea
\delta g_{1in,1out}^{WC} &=& g_{1in,1out} - \nu_{2}\frac{e^{2}}{h}\nonum\\
&=& - c_{1} \frac{e^{2}}{h} \lambda_{1}^{2}T^{2\nu_{2}-2}
+ c_{2} \frac{e^{2}}{h} \lambda_{2}^{2}T^{-2\nu_{2}/\nu_{1}}\nonum\\
\delta g_{1in,2out}^{WC} &=& g_{1in,2out} - (\nu_{1}-\nu_{2})\frac{e^{2}}{h}\nonum\\
&=& + c_{3} \frac{e^{2}}{h} \lambda_{1}^{2}T^{2\nu_{2}-2}
- c_{4} \frac{e^{2}}{h} \lambda_{2}^{2}T^{-2\nu_{2}/\nu_{1}}~,
\eea
where $(c_{1},\ldots,c_{4})$ are non-universal constants. We conclude by noting
that, for the
qparticle-qhole symmetric critical
constriction filling factor of $\nu_{2}^{*}$, the tunneling strengths
$\lambda_{1}\equiv\lambda_{2}$ while the constants are related as
$c_{1}=c_{2}~,~c_{3}=c_{4}$,
such that the corrections to the chiral linear conductances given above vanish,
$\delta g_{1in,1out}^{WC,\nu_{2}^{*}}=0=\delta g_{1in,2out}^{WC,\nu_{2}^{*}}$ .
This provides a more
quantitative explanation of the edge-bias independent constriction transmission observed
experimentally~~\cite{roddaro1}.

\section{Comparison with the experiments}
In this section, we bring together all our results in order to compare them with
the findings of the experiments~~\cite{roddaro1,roddaro2,chung,roddaro3}. In our
phenomenological model for edge state transport in the presence of a gate-voltage
controlled constriction, we have chosen to model the constriction by a mesoscopic
region of lowered electronic density (and hence, filling fraction $\nu_{2}$)
in comparison to that in the bulk (of filling fraction $\nu_{1}$).
The reduced transmission through the constriction at high edge-bias
(i.e., partial transmission ballistic transport) is explained in
terms of the transmitting edge states of the constriction, whose properties are
governed solely by the constriction quantum Hall fluid. Further, the experimentally
observed current backscattered from the constriction (and received in a
terminal on the opposite side of the Hall bar) is explained in terms of the
existence of a gapless edge state lying in between the bulk and constriction
quantum Hall fluids (and whose properties are governed by both the bulk as well
as the constriction fluids).
\par
Next, the startling dip-to-peak evolution of the
vanishing bias constriction transmission with decreasing strength of the
gate-voltage is understood as the result of a competition between
local interedge quasiparticle (quasihole) tunneling events among the transmitting
(reflecting) edges of the constriction region. The edge-bias independent
constriction transmission is seen to arise from the qparticle-qhole symmetric
critical theories that lie on the separatrix of the RG phase diagram; the
equal and opposite growth of the two interedge tunneling processes leads to
an exact cancellation of any gap generating processes. These critical theories
are characterised by a critical value of $\nu_{2}=\nu_{2}^{*}$ and separate
relevant RG flows to either of the cases of perfect and zero transmission through
the constriction. The values of the critical $\nu_{2}^{*}$ and the constriction
transmission at this value, $t(\nu_{2}^{*})$, are seen to match exactly those
obtained experimentally for the cases of the bulk $\nu_{1}=1,1/3$
~~\cite{roddaro1,roddaro2}. Further, the experimental finding of $t(\nu_{2}^{*})$
lying between $0.7$ and $0.8$ for both the cases of $\nu_{1}=2/5$ and
$\nu_{1}=3/7$~~\cite{chung} appear to be consistent with those obtained
from the results obtained from our model. More rigorous experimental investigations
of the $\nu_{1}=2/5,3/7$ systems may, however, be necessary to make a firmer
statement. Further, the two dip-to-peak evolutions observed for the case of
$\nu_{1}=2$~~\cite{roddaro1}, is easily understood from our model as long as we
assume zero inter-Landau level mixing. Finally, the phenomenological assumptions
of (i) a constriction region characterised simply by one parameter, $\nu_{2}$,
and (ii) only local interedge tunneling events inside the constriction appear
to be robust against changes in the size and shape of the constriction (i.e.,
the size and shape of the gate-voltage plates which create the constriction through
electrostatic means); this is consistent with the experimental results 
of Ref.(\cite{roddaro3}).
\par
We now evaluate the relevance of our model in understanding the findings of a
recent work on a constriction circuit with the bulk at
$\nu_{1}=3$~~\cite{miller}. In this work, the authors appear to find
evidence for a zero-bias resistance peak for the case of the constriction
at $\nu_{2}=5/2$. Levin etal., in a very recent work~~\cite{levin}, attribute
this to the possibility of a Pfaffian (Pf) $5/2$ state existing in the constriction
quantum Hall fluid, rather than its particle-hole conjugate anti-Pfaffian (Apf)
state. For this, they appear to use the simple model for the constriction first
revealed in Ref.(\cite{epl}) (and further elaborated upon here), together with
the fact that they find the interedge tunnel coupling between the transmitting edges
($\lambda_{1}$ in our work) to be more RG relevant than the interedge tunnel
coupling between the reflecting edges ($\lambda_{2}$ in our work). Thus, this
experimental observation~~\cite{miller} appears, at first sight, to be
inconsistent with our results
for the case of $\nu_{1}=3$: specifically, our analysis predicts that the constriction
filling fraction of $\nu_{2}=5/2$ is critical, and the constriction transmission
should therefore be edge-bias independent (as observed in the experiments
for the cases of $\nu_{2}=1/2, \nu_{1}=1$ and $\nu_{2}=3/2,
\nu_{1}=2$~~\cite{roddaro1}). While Miller etal.~~\cite{miller} find a zero-bias
resistance peak at $\nu_{2}=7/3$ in consistency with the predictions from our
analysis, experimental results at $\nu_{2}=8/3$ (where our analysis predicts the
existence of a zero-bias resistance minimum) are as yet forthcoming.
\par
A closer look, however, reveals that the inconsistency for the case of $\nu_{2}=5/2$
arises from the following
fact. The analysis of the quantum Hall constriction system carried out in the
present work relies on the assumption that the quantum Hall ground states in
the bulk as well as constriction regions are completely spin polarised, have only
two-body interactions, exactly $1/2$-electron per magnetic flux and no
inter-Landau level mixing. The Pf state, on the other hand, is the exact ground
state of a three-body interaction which explicitly breaks particle-hole
symmetry~~\cite{greiter}. It is, therefore, not surprising that the predictions
of edge state theories based on these two bulk quantum Hall states should be
quite different. A more systematic
experimental study of the $\nu_{1}=3$ constriction system along the
lines of that conducted by Roddaro etal. for the $\nu_{1}=1$ system~~\cite{roddaro1}
could, therefore, improve considerably our understanding of the nature of the
$\nu=5/2$ quantum Hall ground state.

\section{Summary and Outlook}
In this work, we have introduced a model which describes
intermediate conductance scenarios in the problem of tunneling
in 1D chiral systems by constructing a model for a constricted region
(i.e., with a lower filling fraction, $\nu_{2}$, than that of the Hall
fluid in the bulk, $\nu_{1}$). A Landauer-Buttiker analysis of ballistic
transport reveals that the constriction acts as a junction for the
chiral density waves incident on it by splitting them into currents on
the transmitting and reflecting arms of the constriction region.
An edge state model is then formulated
in terms of a hydrodynamic theory of long-wavelength, low-energy
chiral density-wave excitations.
Specifically, we are able to describe the dynamics of the
constriction junction in terms of two pairs of edge fields,
$(\phi^{u},\phi^{d})$ and $(\phi^{l},\phi^{r})$, whose properties are
governed by the effective filling fractions $\nu_{2}$ and
$\nu_{1} - \nu_{2}$ respectively. The constriction is connected to
two incoming- and two outgoing-chiral modes.
Introducing local quasiparticle tunneling across various arm pairs of the
constriction,
we derive the perturbative RG equations for the various tunnel couplings
and find that the RG flow is towards strong-coupling
for both the tunnel couplings considered. A competition between the two
couplings to reach the strong-coupling regime determines the low-energy
configuration of the system. A quasiparticle-quasihole symmetry of the
system is found to determine the existence of a line of critical (gapless)
theories at a critical filling fraction of the constriction region
($\nu_{2}^{*}$) which separates the relevant RG flows towards the perfect
transmission and reflection strong coupling fixed point theories.
The conductances $g_{1in,1out}$ and $g_{1in,2out}$ are computed in the
weak and strong quasiparticle tunneling coupling limits and are also
found to match qualitatively the experimental findings of
Refs.\cite{roddaro1,roddaro2}.
We are also able
to recover the familiar results for $\nu_{2}=\nu_{1}$ \cite{kane,moon}.
In this way, we have achieved a non-trivial generalisation of the generic
phenomenological model of tunneling in FQHE systems formulated by
Kane and Fisher \cite{kane,moon}.
\par
Given the success the phenomenological edge model proposed in this work
meets in providing explanations for the various puzzling experimental
observations, we now turn to a discussion of certain aspects of the
model. First, the model relies essentially
on treating the filling fractions of the quantum Hall ground state in the bulk
($\nu_{1}$) and constriction($\nu_{2}$) regions as the two parameters
of the model. While such a model is sensible for the case of when both
$(\nu_{1},\nu_{2})$ take values from among the special fractions
representing incompressible quantum Hall ground states, how far can we
trust it for the case of when the quantum Hall ground states in the bulk
and constriction regions are compressible? The answer could lie in
a work by Levitov, Shytov and Halperin~~\cite{levitov} provides a
generalisation of the chiral TLL edge state to the case of a compressible
quantum Hall state in the bulk via a composite-fermion Chern-Simons
formulation. Thus, it should be possible to derive the edge model for the
constriction system proposed here for a general quantum Hall ground state
by proceeding along similar lines.
\par
Further, the multimode nature of the edge state of
several members of the incompressible Jain heirarchy of quantum Hall states,
involving one charged edge mode and several charge neutral modes,
is a well established fact theoretically. Thus, it is worth understanding
how the present model (with its assumption of a single edge mode
everywhere in the circuit) fares so well in its comparison with the experiments
even in describing such quantum Hall states. A possible microscopic explanation
can be found by
assuming that the velocity of the charge mode is much greater than
those of the various neutral modes of a multimode edge state~~\cite{leewen}.
Under such circumstances, the dynamics of only the charge edge mode becomes
important in an intermediate energy regime, while the responses arising from
the various neutral modes can be ignored. Further, the composite fermion
field theoretic formulation by Lopez and Fradkin departs from the multimode
picture of the QH edge for the incompressible Jain
fractions~~\cite{lopezfradkin}; this construction involves only one charge
mode (and two auxiliary Klein factors which do not have any additional
propagating degrees of freedom). In this sense, this formulation can be likened
to the other multimode edge theories with vanishing neutral mode velocities.
Finally, for the case of integer quantum Hall systems, it is worth noting
that recent numerical works by Siddiki and co-workers~~\cite{siddiki}
suggest that quantum Hall systems at higher (integer) $\nu$ filling fractions
generically involve only one edge mode in charge transport.
\par
The present work can clearly be generalised to the case of more than
4 chiral wires meeting at a junction. This is important
in modeling the transport across various kinds of junctions of
several non-chiral Tomonaga-Luttinger liquid wires, where the RG phase
diagram is known to contain several non-trivial fixed points
\cite{nayak,lal,chen,chamon,das,sumadip}. Studies of
resonant-tunneling junctions (i.e., junctions which possess a resonant
two-level system or spin-1/2 degree of freedom in addition to allowing
quasiparticle tunneling)
\cite{kane,nayak,chamon,sumadip} reveal interesting transport
phenomena (including variations of the Kondo- and Coulomb-blockade
effects); a resonant-tunneling constriction junction is likely to
possess novel variations of the phenomena found in these works.
\par
Several
experimental studies of noise correlations in quantum Hall edge systems
(both the integer \cite{henny,oberholzer} as well as fractional
\cite{comforti,chung} kind), where Hanbury Brown-Twiss type
correlations have been analysed through current-splitters created using
split-gated constrictions on quantum Hall samples, have thrown up
many interesting results. A similar study has
also been performed with a point contact in a 2DEG \cite{oliver}.
Recently, experimental observations on the interference fringes in the
source-drain conductance of a Mach-Zehnder interferometer made out of
edge states in the integer quantum Hall system have received a lot of
attention~~\cite{ji}.
It would, therefore, be very interesting to study the
current-noise correlations
of the constriction junction model we have set up in the present work.
\par
This study has also revealed the existence of novel gapless edge states
that lie in between gapped quantum Hall fluids with differing filling
factors. In our study, such states carried the reflected current between
the two edges of the Hall bar, making the scenario essentially one of
intermediate transmission. While the phenomenological hydrodynamic edge
state model developed in the present work, and containing these novel
edge states, meets with considerable success in explaining the various
puzzles presented by the experiments \cite{roddaro1,chung}, it will be
even more satisfying to
explore the emergence of such a model from that of a theory containing
the bulk degrees of freedom as well. Such an investigation can be carried
out by starting from a Chern-Simons Ginzburg-Landau type theory \cite{zhang}
of a quantum Hall with a spatially dependent filling factor, and will be the
focus of a future work.
Finally, we note that it remains a challenge to be
able to develop a microscopic understanding of the dependence of the
constriction filling-fraction $\nu_{2}$ proposed in our model on the
gate-voltage $V_{g}$. Accomplishing this will allow us to make detailed
quantitative comparisons with available experimental data as well as
propose future experiments.

\begin{acknowledgments}
I am grateful to D. Sen and S. Rao for many stimulating discussions and
constant encouragement. I would like to thank A.  Altland, B. Rosenow
and Y.Gefen for many invaluable discussions on various aspects of
quantum Hall edge state transport as well as S. Roddaro, V. Pellegrini
and M. Heiblum in helping my understanding of their experiments.
Special thanks are also due to S. Vishveshwara, M. Stone, G. Murthy,
A. Siddiki, R. Mazzarello, F. Franchini, I. Safi, F. Dolcini and A.
Nersesyan for discussions on quantum Hall physics. I am indebted to
Centre for Condensed Matter Theory, Indian Institute of Science,
Bangalore and Harishchandra Research Institute, Allahabad for their
hospitality while this work was carried out.
\end{acknowledgments}

\end{document}